\documentclass[11pt]{article}
\pdfoutput=1

\usepackage{jheppub}
\usepackage{url,comment}
\usepackage{times}
\usepackage{latexsym}
\usepackage{graphicx, graphics, hyperref, amsmath, amssymb, slashed, xcolor, bbm,bm,amsthm, array}
 \usepackage{subfigure}
 \usepackage{listings} 
 \usepackage{trimclip}
\usepackage{rotating}
\usepackage{afterpage}
 \usepackage[colorlinks=true,citecolor=blue,urlcolor=blue]{hyperref}

\newcommand{\nc}{\newcommand}

\nc{\beq}{\begin{equation}}
\nc{\eeq}{\end{equation}}
\nc{\barray}{\begin{eqnarray}}
\nc{\earray}{\end{eqnarray}}
\nc{\barrayn}{\begin{eqnarray*}}
\nc{\earrayn}{\end{eqnarray*}}
\nc{\bcenter}{\begin{center}}
\nc{\ecenter}{\end{center}}
\nc{\mc}{\mathcal}
\nc{\er}[1]{(\ref{eq:#1})}
\nc{\onehalf}{\frac{1}{2}} 
\nc{\partialbar}{\bar{\partial}}
\nc{\psit}{\widetilde{\psi}}
\nc{\Tr}{\mbox{Tr}}
\nc{\hc}{\mbox{H.c.}}
\nc{\ev}{\;\mathrm{eV}}
\nc{\mev}{\;\mathrm{MeV}}
\nc{\gev}{\;\mathrm{GeV}}
\nc{\kev}{\;\mathrm{keV}}
\nc{\tev}{\;\mathrm{TeV}}

\def\chii0{\chi_i^0}
\def\chij0{\chi_j^0}

\newcommand{\gsim}{\lower.7ex\hbox{$\;\stackrel{\textstyle>}{\sim}\;$}}
\newcommand{\lsim}{\lower.7ex\hbox{$\;\stackrel{\textstyle<}{\sim}\;$}}
\nc{\ttbar}{t\bar t}

\def\dbar{{\mathchar'26\mkern-12mu d}}

\def\beq{\begin{equation}}
\def\eeq{\end{equation}}
\def\bea{\begin{eqnarray}}
\def\eea{\end{eqnarray}}

\newcommand{\cref}[1]{Chapter~\ref{c.#1}}

\def\beq{\begin{equation}}
\def\eeq{\end{equation}}
\def\bea{\begin{eqnarray}}
\def\eea{\end{eqnarray}}

\def\GeV{\textrm{GeV}}

\graphicspath{{plots/}}

\title{Cosmological Limits on the Neutrino Mass and Lifetime}

\author[a]{Zackaria Chacko,}
\emailAdd{zchacko@umd.edu}
\author[a]{Abhish Dev,}
\emailAdd{abhishdev92@gmail.com}
\author[a]{Peizhi Du,}
\emailAdd{peizhi.du@stonybrook.edu}
\author[b]{Vivian Poulin}
\emailAdd{vivian.poulin@umontpellier.fr}
\author[a]{and Yuhsin Tsai}
\emailAdd{yhtsai@umd.edu}

\affiliation[a]{Maryland Center for Fundamental Physics, Department of Physics,
University of Maryland, College Park, MD 20742-4111 USA}
\affiliation[b]{Laboratoire Univers \& Particules de Montpellier (LUPM), CNRS \& Universit\'e de Montpellier (UMR-5299),Place Eug\`ene Bataillon, F-34095 Montpellier Cedex 05, France}

\abstract{

At present, the strongest upper limit on $\sum m_{\nu}$, the sum of 
neutrino masses, is from cosmological measurements. However, this bound 
assumes that the neutrinos are stable on cosmological timescales, and is 
not valid if the neutrino lifetime is less than the age of the universe. 
In this paper, we explore the cosmological signals of theories in which 
the neutrinos decay into invisible dark radiation on timescales of order 
the age of the universe, and determine the bound on the sum of neutrino 
masses in this scenario. We focus on the case in which the neutrinos 
decay after becoming non-relativistic. We derive the Boltzmann equations 
that govern the cosmological evolution of density perturbations in the 
case of unstable neutrinos, and solve them numerically to determine the 
effects on the matter power spectrum and lensing of the cosmic microwave 
background. We find that the results admit a simple analytic 
understanding. We then use these results to perform a Monte Carlo 
analysis based on the current data to determine the limit on the sum of 
neutrino masses as a function of the neutrino lifetime. We show that in 
the case of decaying neutrinos, values of $\sum m_{\nu}$ as large as 0.9 
eV are still allowed by the data. Our results have important 
implications for laboratory experiments that have been designed to 
detect neutrino masses, such as KATRIN and KamLAND-ZEN.
 
}

\begin{document}

\begin{flushright}
\small{.}
\end{flushright}

\maketitle

\section{Introduction}\label{s.introduction}

Over the last few decades, a series of oscillation experiments have 
convincingly established that the neutrinos have masses, and determined 
their mass splittings. However, the actual values of the masses of the 
three neutrino species continue to remain a mystery. In particular, it 
is still not known whether the spectrum of neutrino masses is 
hierarchical, inverse hierarchical or quasi-degenerate. The question of 
whether the neutrino masses are Dirac or Majorana also remains 
unanswered.

At present, the strongest limit on the sum of neutrino masses, $\sum 
m_\nu < 0.12$ eV, is from cosmological 
observations~\cite{Aghanim:2018eyx}. These measurements are sensitive to 
the neutrino masses through the gravitational effects of the relic 
neutrinos left over from the Big Bang. In determining the size of this 
effect~\cite{Bond:1980ha,Hu:1997mj}, (reviews with additional references 
may be found in 
~\cite{Wong:2011ip,Lattanzi:2017ubx,Lesgourgues:2018ncw,Tanabashi:2018oca}), 
it is assumed that the neutrinos are stable on timescales of order the 
age of the universe. In particular, if the neutrino lifetime is less 
than the age of the universe~\cite{Serpico:2007pt,Serpico:2008zza}, or 
if the neutrinos have annihilated away into lighter 
states~\cite{Beacom:2004yd,Farzan:2015pca}, this bound on the neutrino 
masses is no longer valid and must be reconsidered. In this paper, we 
explore the cosmological signals that arise from a general framework in 
which the neutrinos decay into dark radiation on timescales shorter than 
the age of the universe, and determine the bound on the sum of neutrino 
masses as a function of the neutrino lifetime in this scenario. Our 
focus is on the case in which neutrinos decay after becoming 
non-relativistic.

The case for neutrino decay is theoretically extremely well-motivated. 
Neutrino decay is in fact a characteristic feature of models in which 
neutrinos have masses. Even in the minimal extensions of the Standard 
Model (SM) that incorporate Dirac neutrino masses by adding right-handed 
neutrinos, or Majorana masses by including the non-renormalizable 
Weinberg operator, the heavier neutrinos undergo two-body decays at one 
loop into a lighter neutrino and a 
photon~\cite{Petcov:1976ff,Goldman:1977jx,Marciano:1977wx,Lee:1977tib,Pal:1981rm}, 
(useful discussions may also be found 
in~\cite{Mohapatra:1991ng,1992pmn..book.....B}). In these scenarios, the 
lifetime of the massive neutrino is given by $\tau_{\nu}\sim 
10^{50}\textrm{s}\left({0.05\,\textrm{eV}}/{m_{\nu}}\right)^5$, assuming 
the daughter neutrino mass is negligible. This is much longer than the 
age of the universe, and therefore these minimal frameworks do not give 
rise to observable cosmological signals from neutrino decay. However, in 
more general extensions of the SM that incorporate neutrino masses, the 
neutrino lifetime can be much shorter. In particular, this includes 
theories in which the generation of neutrino masses is associated with 
the spontaneous breaking of global symmetries in the neutrino 
sector~\cite{Gelmini:1980re,Chikashige_1981,Georgi:1981pg,VALLE198387,Gelmini:1983ea}, 
(see also~\cite{Dvali:2016uhn,Funcke:2019grs}). In this framework, the 
heavier neutrinos can decay into a lighter neutrino and one of the 
Goldstone bosons associated with the spontaneous breaking of the global 
symmetry. The timescale for this process can be shorter than or 
comparable to the age of the universe, giving rise to cosmological 
signals. In general, neutrinos that are unstable on cosmological 
timescales remain an intriguing possibility due to the strong 
motivations for new physics that explains the smallness of neutrino 
masses.

In the past, the decaying neutrino scenario has been explored as a 
solution to the solar and atmospheric neutrino 
problems~\cite{Bahcall:1972my,Berezhiani:1991vk,Acker:1991ej,Barger:1998xk}. 
However, the resulting predictions for the energy spectrum of the solar 
neutrinos and the decay lengths required for this proposal are now 
disfavored by the 
data~\cite{Acker:1993sz,CHOUBEY200073,Joshipura:2002fb}. There has also 
been earlier work studying the impact of the decay of massive neutrinos 
on structure formation~\cite{Doroshkevich:1984,Doroshkevich:1988}. 
However the range of parameter space that was considered is much above 
the current limits on the masses of the neutrinos. More recently, 
radiative neutrino decays have been proposed as an explanation of the 21 
cm signal observed by the EDGES experiment~\cite{Chianese:2018luo}.

The current limits on the neutrino lifetime are rather weak, except in 
the case of decays to final states involving photons. In this specific 
case, the absence of spectral distortions in the cosmic microwave 
background (CMB) places strong bounds on radiative decays from a heavier 
neutrino mass eigenstate to a lighter one, $\tau_{\nu}\gsim 10^{19}$s 
for the larger mass splitting and $\tau_{\nu}\gsim 4 \times 10^{21}$s 
for the smaller one~\cite{Aalberts:2018obr}. There are also very strong, 
albeit indirect, limits on radiative neutrino decays based on the tight 
laboratory and astrophysical bounds on the neutrino dipole moment 
operators that induce this 
process~\cite{Beda:2013mta,Borexino:2017fbd,Raffelt:1990pj,Raffelt:1999gv,Arceo-Diaz:2015pva}.

In contrast, the decay of neutrinos into dark radiation that does not 
possess electromagnetic interactions is only weakly constrained by 
current cosmological, astrophysical, and terrestrial data. The most 
stringent bound on this scenario arises from CMB measurements. If 
neutrino decay and inverse decay processes are effective during the CMB 
epoch, they prevent the neutrinos from free streaming, leading to 
observable effects on the CMB~\cite{Peebles,Hu:1995en,Bashinsky:2003tk}. 
Current measurements of the CMB power spectra require neutrinos to free 
stream from redshifts $z \approx 8000$ until recombination, $z \approx 
1100$ 
\cite{Archidiacono:2013dua,Audren:2014lsa,Follin:2015hya,Escudero:2019gfk}.\footnote{Also 
see the more recent discussion in \cite{Kreisch:2019yzn} for the effects 
of interacting neutrinos on the CMB.} This can be used to set a lower 
bound on the neutrino lifetime $\tau_\nu\geq 4\times 
10^{8}\,\textrm{s}\left({m_{\nu}}/{0.05\,\textrm{eV}}\right)^3$ for SM 
neutrinos decaying into massless dark radiation~\cite{Escudero:2019gfk}. 
Several astrophysical observations have also been used to set limits on 
the neutrino lifetime. However, the resulting bounds are much weaker. 
The observation that the neutrinos emitted by Supernova 1987A did not 
decay prior to reaching the earth can be used to set a bound on the 
lifetime of the electron-neutrino, $\tau_{\nu_e}/m_{\nu_e}\geq 5.7\times 
10^5\, \textrm{s}/\textrm{eV}$~\cite{Frieman:1987as}. Similarly, the 
detection of solar neutrinos at the earth can be used to place a bound 
on the lifetime of the mass eigenstate $\nu_2$, $\tau_{\nu}/m_\nu\gtrsim 
10^{-4}\, \textrm{s}/\textrm{eV}$ 
\cite{Joshipura:2002fb,Beacom:2002cb,Bandyopadhyay:2002qg}. Limits on 
the neutrino lifetime can also be obtained from atmospheric neutrinos 
and long-baseline experiments, but the resulting constraints are even 
weaker (see e.g. 
\cite{GonzalezGarcia:2008ru,Gomes:2014yua,Choubey:2018cfz,Aharmim:2018fme}). 
Therefore, at present there is no evidence that neutrinos are stable on 
cosmological timescales, and that the cosmic neutrino background 
(C$\nu$B) has not decayed away into dark radiation.

The impact of non-vanishing neutrino masses on cosmological structure 
formation is well understood, (see 
\cite{Wong:2011ip,Lesgourgues:2018ncw} for useful reviews).
 \begin{itemize}
 \item 
 Sub-eV neutrinos constitute radiation at the time of matter-radiation 
equality. Therefore, fluctuations about the background neutrino number 
density do not contribute significantly to the growth of structure until 
after neutrinos have become non-relativistic. Consequently, 
perturbations on scales that enter the horizon prior to neutrinos 
becoming non-relativistic evolve differently than scales that enter 
afterwards, thereby affecting the matter power spectrum.
 \item
 After neutrinos become non-relativistic, their overall contribution to 
the energy density redshifts away less slowly than that of a 
relativistic species of the same abundance. This results in a larger 
Hubble expansion, reducing the time available for structure formation. 
This leads to an overall suppression of large scale structure (LSS).
 \end{itemize}
 Then the leading effect of non-vanishing neutrino masses is to suppress 
the growth of structure on scales that entered the horizon prior to the 
neutrinos becoming non-relativistic. The extent of this suppression 
depends on the values of the neutrino masses. Since heavier neutrinos 
become non-relativistic earlier and also contribute a greater fraction 
of the total energy density after becoming non-relativistic, a larger 
neutrino mass results leads to more suppression of LSS. In the case of 
neutrinos that decay, this suppression now also depends on the neutrino 
lifetime. After neutrinos have decayed, their contribution to the energy 
density redshifts like that of massless neutrinos, resulting in a milder 
suppression of structure as compared to stable neutrinos of the same 
mass. It follows that there is a strong degeneracy between the neutrino 
mass and the lifetime inferred from the matter power spectrum. The 
cosmological upper bound on the neutrino mass is therefore 
lifetime-dependent, as was first discussed 
in~\cite{Serpico:2007pt,Serpico:2008zza}.

Neutrino masses also lead to observable effects on the CMB. Sub-eV 
neutrinos become non-relativistic after CMB decoupling. The main 
``primary'' effect on the CMB is through the early and late 
integrated-Sachs-Wolfe effects, as well as a modification of the angular 
diameter distance to the last scattering surface. Because of their 
impact on the growth of structure detailed above, neutrinos also affect 
the CMB through the ``secondary'' effect of lensing. At the precision of 
{\em Planck}, the effects of lensing drive the CMB constraints on the 
sum of neutrino masses. Since neutrino decay results in a milder 
suppression of structure as compared to stable neutrinos of the same 
mass, the bounds on $\sum m_\nu$ from CMB lensing are also lifetime 
dependent.

We begin our analysis by deriving the Boltzmann equations that govern 
the cosmological evolution of density perturbations in the case of 
unstable neutrinos. We then appropriately modify the Boltzmann code {\sf 
CLASS}\footnote{http://www.class-code.net} \cite{Blas:2011rf} to 
calculate the CMB and matter power spectra to accommodate this 
framework. We find that the results admit a simple analytic 
understanding. We then perform a Monte Carlo analysis based on CMB and 
LSS data (Planck+BAO+Pantheon+LSS) to determine the bounds on this 
scenario. We use the likelihood function from the Planck 2015 
analysis~\cite{Ade:2015xua}.{\footnote{While this analysis was being 
finalized, the Planck 2018 data became public~\cite{Aghanim:2019ame}. We leave the analysis using Planck 2018 data to future work.
} We find that when the stable neutrino 
assumption is relaxed, the limits on the neutrino masses from this data 
set become much weaker, with the bound on $\sum m_\nu$ increasing from 
0.25 eV to 0.9 eV. Importantly, this shows that the cosmological bounds 
do not exclude the region of parameter space in which future experiments 
such as KATRIN~\cite{Angrik:2005ep}, KamLAND-ZEN 
(KLZ)~\cite{KamLAND-Zen:2016pfg} and the Enriched Xenon Observatory 
(EXO)~\cite{Auger:2012gs,Albert:2014awa} are sensitive to the neutrino 
masses.

Our focus in this paper is on the decay of neutrinos to dark radiation, 
since this framework has a greater impact on the bound on $\sum m_\nu$ 
than the decay of heavier neutrinos to lighter ones. In particular, at 
present the cosmological limits on $\sum m_\nu$ only constrain 
quasi-degenerate neutrino spectra, so that decays of heavier neutrinos 
to lighter ones are not expected to alter the current bound 
significantly. In appendix~\ref{app:1} we present an example of a simple 
model in which the neutrinos decay into dark radiation on timescales of 
order the age of the universe. This model is consistent with all current 
cosmological, astrophysical and laboratory bounds, and represents a 
concrete realization of the scenario we are considering. However, we 
stress that the results presented in the body of the paper are not 
restricted to this specific model, but apply to any theory in which the 
neutrinos decay to dark radiation after becoming nonrelativistic.

The outline of this paper is as follows. In the next section we discuss 
the parameter space of the neutrino mass and lifetime, outlining the 
current bounds. In Sec.~\ref{sec:numerics}, we derive the Boltzmann 
equations that dictate the cosmological evolution of perturbations in 
the phase-space distribution of unstable neutrinos and their daughter 
radiation. While our focus is on the case in which the decaying 
particles are neutrinos, the formalism is more general and can be 
applied to the much larger class of models in which warm dark matter 
decays into dark radiation. In Sec.~\ref{sec:signature}, we numerically 
compute the growth of perturbations in the case of unstable neutrinos, 
and determine the effects on the matter power spectrum and on CMB 
lensing. To obtain a physical understanding, in 
Sec.~\ref{sec:analytical} we derive analytical expressions for these 
effects. In Sec.~\ref{sec:MCstudy}, we perform a Monte Carlo scan of the 
parameter space and derive constraints on the mass and lifetime of the 
neutrino from current data. Our conclusions are in 
Sec.~\ref{sec:conclusion}. In appendix~\ref{app:1}, we present a 
realistic example of a model in which the neutrinos decay into dark 
radiation on timescales of order the age of the universe.

\section{Parameter Space of the Unstable Neutrino}\label{sec:param}
\begin{figure}
\begin{center}
\includegraphics[scale=0.4]{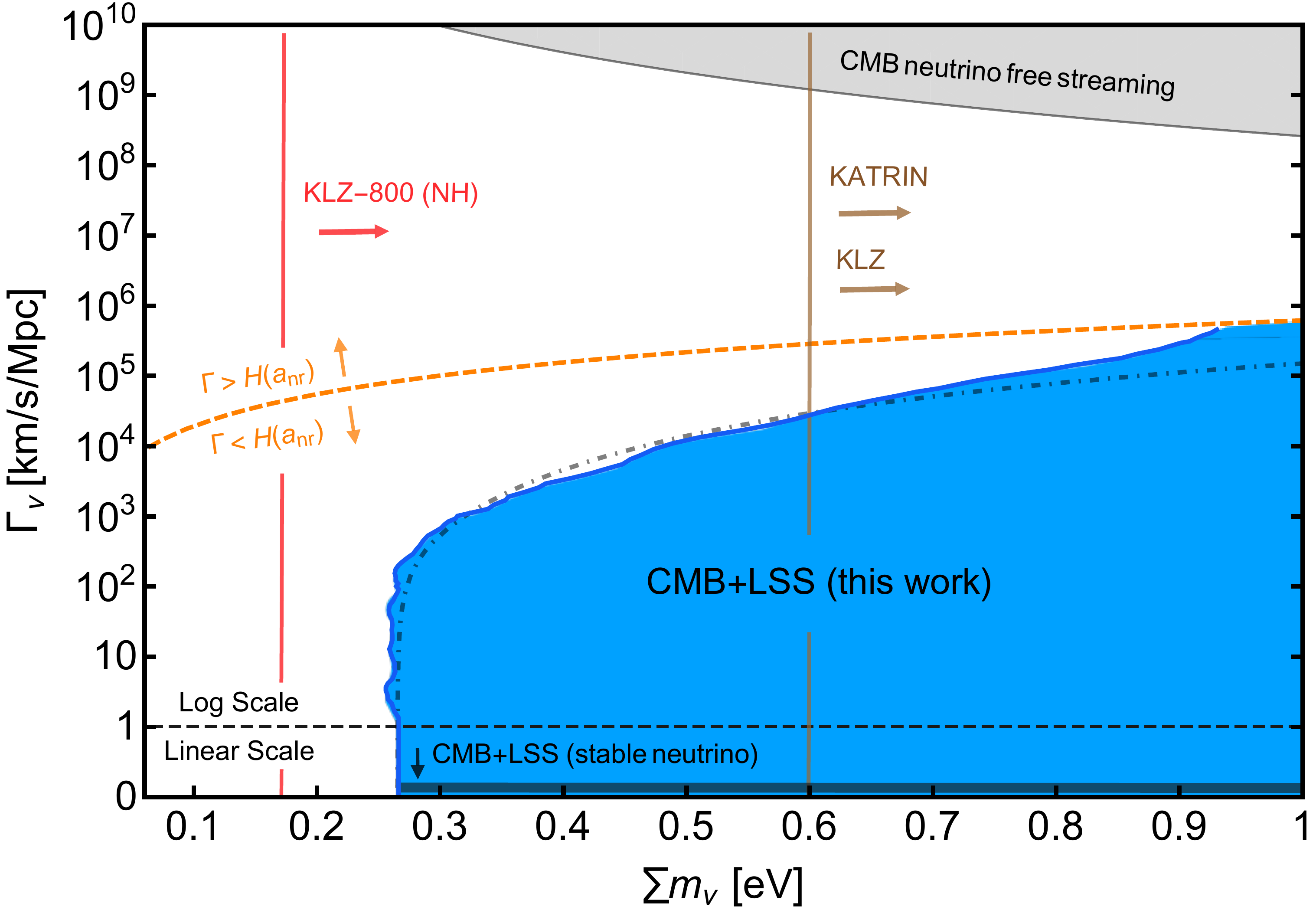}
 \caption{The plot shows the current constraints in the $\sum 
m_\nu-\Gamma_\nu$ parameter space. The colored regions are excluded by 
current data while the white region is allowed. The orange dashed line 
separates the region of parameter space in which neutrinos decay while 
still relativistic from that in which they decay after becoming 
non-relativistic. Our study focuses on the region below this line, 
corresponding to the latter scenario. The light grey regions show 
current constraints on neutrino mass and lifetime coming from CMB free 
streaming and the bound on stable neutrinos (labelled ``CMB+LSS (stable 
neutrino)''). Our analysis excludes the blue region labelled ``CMB+LSS 
(this work)'' based on CMB and LSS data (Planck+BAO+Pantheon+LSS). The 
dash-dotted line represents the approximate constraint obtained by 
simply requiring that the matter power spectrum be consistent with 
observations in the neighborhood of $k = 0.1\,h$/Mpc with fixed $H_0$. 
This is seen to provide a reasonable estimate to the constraints from 
all data. The vertical brown band shows the projected KATRIN sensitivity 
and also the current KLZ sensitivity. The vertical red line shows the 
projected KLZ-800 sensitivity in the case of a normal hierarchy. }
 \label{TargetSpace}
 \end{center}
 \end{figure}

In this section we outline the constraints on the decay of neutrinos to 
dark radiation. As explained in the introduction, these bounds only 
place limits on a combination of the neutrino mass and the lifetime. 
Therefore, in this study we will map out the constraints and the signals 
in the two-dimensional parameter space spanned by the sum of neutrino 
masses ($\sum m_\nu$) and the neutrino decay width ($\Gamma_\nu$), as 
displayed in Fig.~\ref{TargetSpace}. In our analysis we make the 
simplifying assumption that all three neutrinos are degenerate in mass. 
As we shall see, the bounds on $\sum m_\nu$ are always much larger than 
the observed mass splittings, and so this is an excellent approximation 
in the relevant parameter space. We further assume that all three 
neutrinos have the same decay width $\Gamma_{\nu}$. Since the mixing 
angles in the neutrino sector are large, this is a good approximation in 
many simple models of decaying neutrinos if the spectrum of neutrinos is 
quasi-degenerate. In particular, the model presented in 
appendix~\ref{app:1} exhibits this feature.

There is a hard lower limit on the sum of neutrino masses from the 
atmospheric and solar mass splittings which constrain $\sum m_\nu\geq 
\sqrt{\Delta m^2_{31}} + \sqrt{\Delta m^2_{21}} =0.06$ eV in the case of 
normal ordering and $\sum m_\nu\geq 2\times\sqrt{\Delta m^2_{31}}=0.1$ 
eV in the case of inverted ordering \cite{Tanabashi:2018oca}. Therefore, we present the 
parameter space starting from $\sum m_\nu=0.06\,$ eV. CMB observations 
can be used to obtain an upper bound on the sum of neutrino masses. The 
current CMB data constrains the effective number of neutrinos, $N_{\rm 
eff}$, during the epoch of acoustic oscillations to be $2.99\pm 
0.17$~\cite{Aghanim:2018eyx}, which is perfectly compatible with the 
SM value of 3.046. Then, if neutrinos are stable on CMB 
timescales, we can obtain an approximate upper bound on their masses by 
requiring that all three species of neutrinos are relativistic at 
recombination. This translates into an approximate limit, $\sum m_\nu 
\lesssim 3 T_{\rm rec} \approx 0.9$ eV. A more precise bound can be 
obtained from a fit to the CMB data.

The CMB can also be used to constrain the masses of neutrinos that decay 
prior to recombination. As mentioned in the introduction, CMB data 
requires the species that constitute $N_{\rm eff}$ to be free streaming 
at redshifts below $z \approx~8000$ 
until recombination, $z \approx 1100$. 
This can be used to place limits on processes such as neutrino decays 
and inverse decays that prevent neutrinos from free streaming at late 
times. The resulting bound depends on the neutrino mass, and is given by 
$\tau_\nu \geq 4 \times 10^{8}\,\textrm{s} 
\left({m_\nu}/{0.05\,\textrm{eV}}\right)^3$~\cite{Escudero:2019gfk}. 
This bound excludes the grey region at the top of 
Fig.~\ref{TargetSpace}. Naively, one might expect the CMB bounds from 
free streaming to rule out all theories in which the neutrino decays 
before recombination, independent of the neutrino mass. However, in the 
case of an ultrarelativistic mother particle, the decay process results 
in approximately collinear daughter particles moving in the same 
direction as the mother. Similarly the inverse decay process generally 
only involves collinear initial state particles, so that there is no 
significant disruption in the flow of energy even if the decay and 
inverse decay processes are efficient~\cite{Archidiacono:2013dua}. The 
net constraint from CMB free streaming is therefore much weaker on the 
decays of light neutrinos.

As discussed in the introduction, massive neutrinos suppress the growth 
of matter perturbations by reducing the time available for structure 
formation. In the case of stable neutrinos, this has been used to set a 
constraint on the sum of neutrino masses, $\sum m_\nu \leq 0.12$ 
eV~\cite{Aghanim:2018eyx}. Unstable neutrinos that decay after becoming 
non-relativistic also lead to a suppression in the growth of structure 
that now depends on the neutrino lifetime. In this paper we determine 
the resulting bound in the two dimensional parameter space spanned by 
$\sum m_\nu$ and the neutrino lifetime. Based on the Monte Carlo study 
presented in Sec.~\ref{sec:MCstudy}, CMB and LSS data (Planck+BAO+Pantheon+LSS) exclude the blue region labelled as ``CMB+LSS (this 
work)" in Fig.~\ref{TargetSpace}. We have scanned the region between 
$0\leq{\rm log}_{10}\frac{\Gamma_\nu}{\textrm{km/s/Mpc}}\leq5.5$. In 
Fig.~\ref{TargetSpace}, we simply extrapolate the bound at ${\rm 
log}_{10}\frac{\Gamma_\nu}{\textrm{km/s/Mpc}}=0$ to $\Gamma_\nu=0$, 
because the constraint on $\sum m_\nu$ is independent of $\Gamma_\nu$ 
when $\Gamma_\nu\ll H_0$. The existing constraint on the masses of 
stable neutrinos from this data set forms the lower boundary of this 
region (labelled as ``CMB+LSS (stable neutrino)'').

The dash-dotted line that approximately envelopes the blue shaded region 
represents the constraint obtained by simply requiring that the matter 
power spectrum be consistent with observations in the neighborhood of $k 
= 0.1\,h$/Mpc with fixed $H_0$, where the current LSS measurements have 
the best sensitivity. We see that it provides a good approximation 
to the true bound, except in the region of $\sum m_\nu \gtrsim 0.9$ eV, 
where the CMB limits on $N_{\rm eff}$ at recombination become important. 
The impact of neutrinos on the matter power spectrum depends slightly on 
the mass ordering as the individual mass eigenstates become 
non-relativistic at different times. However, since the current limits 
are only sensitive to quasi-degenerate spectra, we are justified in 
neglecting this effect.

The orange dashed line ($\Gamma=H(z_{\rm nr})$) separates the region 
where neutrinos decay when non-relativistic from the region where they 
decay while still relativistic. Here $z_{\rm nr}$, the approximate 
redshift at which neutrinos become non-relativistic, is defined 
implicitly from the relation $3 T_\nu(z_{\rm nr})=m_\nu$. This 
definition is based on the fact that for relativistic neutrinos at 
temperature $T_\nu$, the average energy per neutrino is approximately $3 
T_\nu$. The Hubble scale at $z_{\rm nr}$ is given by, 
  \begin{eqnarray}
      H(z_{\rm nr}) &=& H_0\sqrt{\Omega_m}\bigg(\frac{\sum m_\nu}{9T_\nu^0}\bigg)^{3/2} 
\\
&\simeq&  7.5\times 10^5 {\rm km/s/Mpc} \bigg(\frac{H_0}{68 {\rm km/s/Mpc}}\bigg)
\bigg(\frac{\Omega_m}{0.3}\bigg)^{1/2}\bigg(\frac{\sum m_\nu}{1 {\rm eV}}\bigg)^{3/2}\bigg(\frac{1.5\times 10^{-4} {\rm eV}}{T_\nu^0}\bigg)^{3/2}.\nonumber
  \end{eqnarray}
 Since our study assumes neutrinos decay after they become 
non-relativistic, we only present the constraints below this orange 
dashed line.

The currently allowed parameter space is represented by the white 
regions in Fig.~\ref{TargetSpace}.  In the white region above the orange 
dashed line, even though neutrinos decay when still relativistic, their 
small mass allows them to evade the current CMB free streaming 
constraints. In this scenario their contribution to the energy density 
evolves in a manner similar to that of massless neutrinos, and so the 
effects on LSS are similar in the two cases. In the white region below 
the orange dashed line the neutrinos decay after becoming 
non-relativistic, but because their masses are too small or their 
lifetimes too short, the suppression of the matter power spectrum is too 
small to be detected with current data.

We see from this discussion that the unstable neutrino paradigm greatly 
expands the range of neutrino masses allowed by current data. This has 
important implications for current and future laboratory experiments 
designed to detect neutrino masses. Next generation tritium decay 
experiments such as KATRIN~\cite{Angrik:2005ep} are expected to be 
sensitive to values of $m_{\nu_e}$ as low as $0.2$ eV, corresponding to 
$\sum m_{\nu}$ of order 0.6 eV. A signal in these experiments would 
conflict with the current cosmological bound, $\sum m_{\nu} < 0.12$ eV, 
for stable neutrinos. However, in the decaying neutrino paradigm, we 
have seen that the current cosmological upper bound on the sum of 
neutrino masses is relaxed, with the result that $\sum m_{\nu}$ as high 
as 0.9 eV is still allowed. Therefore, a signal at KATRIN can be 
accommodated if neutrinos are unstable on cosmological timescales. In 
Fig.~\ref{TargetSpace}, we display a brown vertical line $\sum m_{\nu} 
\approx 0.6$ eV that corresponds to the expected KATRIN sensitivity.

In the case of Majorana neutrinos, current data from neutrinoless 
double-beta decay experiments such as KLZ and EXO have already ruled out $\sum 
m_\nu \gtrsim 0.6$ eV (brown vertical 
line)~\cite{KamLAND-Zen:2016pfg,Albert:2014awa}. An updated version of 
KLZ, the KLZ-800, is currently probing $\sum m_{\nu}$ as low as $0.17$ 
eV~\cite{Brunner:2017iql} (red vertical line) in the case of the normal 
hierarchy and the entire parameter space for the inverted hierarchy. If 
this experiment were to see a signal, we cannot immediately conclude 
that hierarchy is inverted based on the current cosmological bound of 
$\sum m_{\nu} < 0.12$ eV, since the decaying neutrino paradigm would 
still admit a normal hierarchy.

\section{Evolution of Perturbations in the Decay of Non-Relativistic 
Particles into Radiation}\label{sec:numerics}

In this section we derive the set of Boltzmann equations describing the 
evolution of the phase-space density of massive particles decaying into 
massless daughter particles, working to first order in the 
perturbations. In contrast to the case of cold dark matter (CDM) decay 
(see, e.g., \cite{Audren:2014bca,Poulin:2016nat}), we cannot assume that 
the mother particles are at rest, but must take into account their 
non-trivial momentum distribution, as in the 
studies~\cite{Kawasaki:1992kg, Bharadwaj:1997dz, Kaplinghat:1999xy}. 
This allows us to study the cosmological effects of a warm particle 
species, such as neutrinos or warm dark matter, decaying into radiation. 
We implement these new Boltzmann equations into the numerical code {\sf 
CLASS} to generate the results in sections~\ref{sec:signature} and 
\ref{sec:MCstudy}.

 The phase-space distribution of a particle species in the expanding 
universe is a function of the position $\vec{x}$, the comoving momentum 
$\vec{q}\equiv q\hat{n}$, and the comoving time $\tau$. The evolution of 
this distribution is determined by the Boltzmann equation,
 \begin{equation}\label{BE}
\frac{df}{d\tau}=\frac{\partial f}{\partial \tau}+\frac{dx^i}{d\tau}\frac{\partial f}{\partial x^i}+\frac{dq}{d\tau}\frac{\partial f}{\partial q}+\frac{d\hat{n}}{d\tau}.\frac{\partial f}{\partial \hat{n}}=C[f] \;,
 \end{equation}
 where $C[f]$ is the collision term that accounts for all processes 
involving the species.

 We consider the case of a massive mother (with the subscript $M$ for 
mother) of mass $M$ decaying into $N$ daughters ($D_{i=1,2...N}$). For 
the sake of simplicity, we restrict ourselves to the case where the 
mother particles decay after becoming non-relativistic, but nevertheless 
keep track of their non-trivial momentum distribution. In this regime, 
inverse-decay processes can be safely neglected. We also ignore any 
effects arising from Pauli blocking and spontaneous emission since 
$f_{M,Di}\ll 1$. The collision terms for the mother and daughter 
particles are then given by,
 \begin{eqnarray}
C_M=&-&\frac{a^2}{2\epsilon_{M}}\int \prod_i\frac{\dbar^3 \vec {q_i}}{2\epsilon_{Di}}|\mathcal{M}|^2(2\pi)^4 \delta^{(4)}(\vec q_M-\Sigma_i \vec q_{Di})f_M(q_M),\label{eq:CM}\\
C_{Dj}=&+&\frac{a^2}{2\epsilon_{Dj}}\int \frac{\dbar^3 \vec {q}_M}{2\epsilon_M}\prod_{i\neq j}\frac{\dbar^3 \vec {q_i}}{2\epsilon_{Di}}|\mathcal{M}|^2(2\pi)^4 \delta^{(4)}(\vec q_M-\Sigma_i \vec q_{Di}) f_M(q_M).
 \end{eqnarray} 
 where $\epsilon_S \equiv(q_S^2+m_S^2a^2)^{1/2}$ represents the comoving 
energy of the species $S( \equiv M, D_i)$ and $\dbar^3 \vec q\equiv d^3 \vec q/(2\pi)^3$.  From the definition of the 
decay width, the collision term for the mother particle can be 
simplified to
 \begin{equation}\label{eq:C_M}
C_M =-\frac{a \Gamma}{\gamma} f_M,
\end{equation}
 where $\Gamma$ denotes the decay width in the rest frame of the 
decaying particle, and the relativistic boost factor $\gamma\equiv 
\sqrt{q_M^2 + M^2 a^2}/(Ma)$ accounts for time-dilation in the cosmic 
frame. To determine the evolution of inhomogeneities in our universe, we 
consider perturbations about the homogeneous and isotropic background 
phase space distribution functions,
 \begin{eqnarray}\label{fperturb}
f_{S}(q_{S}, \hat{n}, \vec{x}, \tau) &=& f^0_{S}(q_{S},\tau)+\Delta f_{S}(q_{S}, \hat{n}, \vec{x}, \tau),\quad S=M,D_i.
 \end{eqnarray}

\subsection{Background: Zeroth Order}
 
 Treating $\Delta f_M$ and fluctuations about the 
gravitational background as higher order perturbations, the zeroth order 
Boltzmann equations for $f^0_M$ arising from 
Eq.~(\ref{BE}) take the form,
 \begin{eqnarray}
\frac{\partial f_M^0}{\partial \tau}&=& -a\frac{\Gamma}{\gamma}f_M^0.\label{DE} 
 \end{eqnarray} 
 The formal solution to $f_M^0(q,\tau)$ from the differential equations 
in Eq.~(\ref{DE}) is given by,
 \bea \label{formal}
f_M^0(q,\tau)=f_{i}(q)e^{-\Gamma\int_{\tau_i}^\tau \frac{ 
a}{\gamma(a)} d\tau'}, 
 \eea 
 where $\tau_i$ denotes the initial conformal time and $f_i(q)$ 
represents the initial momentum distribution. We will focus on the case 
where the mother decays after becoming non-relativistic. Using 
integration by parts, the exponent in Eq.~(\ref{formal}) can be 
rewritten as,
 \bea 
 \label{eq:exponent}
\Gamma \int_{\tau_i}^{\tau}\frac{a\,d\tau'}{\gamma(a)}=\frac{\Gamma t'}{\gamma(a)}\Bigg|_{t_i}^{t}-\Gamma\int_{t_i}^t dt'~ t'\frac{d}{dt'}\left(\frac{1}{\gamma(a)}\right),
 \eea
 where we have used $ad\tau=dt$. It is computationally demanding to 
solve the integral for general $a(\tau)$. However, the behavior of the 
exponential factor is rather simple: the exponential is close to $1$ 
when $\tau$ is smaller than the mother lifetime $\sim\gamma/\Gamma a$, 
and $f_M$ no longer contributes when $\tau$ is much larger than the 
mother lifetime. The only time that the exponential factor exhibits a 
non-trivial $a$-dependence is when $\tau\sim\gamma/\Gamma a$. Since our 
focus is on decays in the non-relativistic regime, so that $\gamma(a)$ 
is slowly varying at the time of decay. Then the second term on the 
right-hand side of Eq.~(\ref{eq:exponent}), which depends on the time 
derivative of $\gamma(a)$, can be neglected in favor of the first term. 
This allows us to approximate the exponent as
 \bea \label{eq:approx_f_M}
\Gamma \int_{\tau_i}^{\tau}\frac{a\,d\tau'}{\gamma(a)}\approx \frac{\Gamma t}{\gamma(a)}.
\eea
 We have verified numerically that Eq.~(\ref{eq:approx_f_M}) is a good 
approximation to the full solution. Therefore, the mother distribution 
we use in this paper is
 \bea\label{eq:f_M^0_sol} 
f_M^0(q,\tau)\approx f_i(q)e^{-\frac{\Gamma}{\gamma(a)} t}.
 \eea
 It is worth pointing out that the mother distribution described by 
Eq.~(\ref{eq:f_M^0_sol}) is a general formula that can also be applied 
to the case of decaying CDM. This limiting case 
corresponds to the distribution $f_i(q_M)=\delta(q_M)N_{Mi}/(4\pi 
q^2_M)$, where $N_{Mi}$ represents the initial comoving number density 
of mother particles. Since this distribution is localized entirely at 
$q_M=0$, the boost factor $\gamma(a)=1$. Then Eq.~(\ref{eq:f_M^0_sol}) 
reduces to the known result for decaying cold dark 
matter~\cite{Aoyama:2011ba, Wang:2012eka, Aoyama:2014tga}. Our analysis 
is, however, more general, because it accounts for the fact that the 
contribution of warm dark matter to the background energy density scales 
with the redshift in a more complicated manner than $a^{-3}$. In 
addition, it takes into account the fact that, in general, particles 
with larger momenta live longer as a consequence of time dilation.

 We now apply the above general formula to the decay of massive 
neutrinos. The SM neutrinos decoupled from the photon bath 
when they were ultra-relativistic. Therefore, their distribution prior to 
decay is of the Fermi-Dirac form. Therefore, $f_i=1/(e^{q/T_{\nu 
0}}+1)$, leading to
 \begin{equation}\label{Mpsd}
f_{M}^0 = \frac{1}{e^{q/T_{\nu0 }}+1}\frac{}{}\exp(-\frac{\Gamma }{\gamma}t).
 \end{equation} 
  The collision terms for the daughter particles are more challenging. 
However, we can simplify this set of equations by using the total 
integrated Boltzmann equations for the daughters. This is done by 
integrating the Boltzmann equations for the individual daughter species 
with respect to $\dbar^3\vec q_{Di} \epsilon_{Di}$ and adding them up. 
The resulting total integrated collision term for the daughter species 
is given by,
 \begin{eqnarray}
    \sum_{j}\int \dbar^3 \vec q_{Dj}\epsilon_{Dj}C^0_{Dj} &=& a^2 \int \frac{\dbar^3 \vec q_M}{2\epsilon_M}\prod_i\frac{\dbar^3\vec q_{Di}}{2\epsilon_{Di}} (\sum_j \epsilon_{Dj})|\mathcal{M}|^2(2\pi)^4 \delta^{(4)}(\vec q_M-\Sigma_i \vec q_{Di}) f^0_M(q_M),\nonumber
    \\
    &=&a^2\Gamma M\int \dbar^3\vec q_M f^0_M.\label{EM}
 \end{eqnarray}
 The simplification in the last line follows from the covariant 
conservation of the energy-momentum tensor, where we have used 
Eq.~(\ref{eq:CM}), Eq.~(\ref{eq:C_M}), and $\epsilon_M/\gamma=Ma$ to 
obtain this expression. In this work we focus on the case in which the 
mother neutrino decays into massless daughter particles. The relation in 
Eq.~(\ref{EM}) can be used to express the Boltzmann equation for the 
daughters in terms of the total comoving energy density of the daughters 
$E_{D}$ and the comoving number density of the mother $N_M$
 \begin{equation}\label{EDNM}
    E_D\equiv  \sum_i \int dq_{D_i}q_{D_i}^3\,f_{D_i},\qquad N_M\equiv \int dq_M q_M^2\, f_M\,.
\end{equation} 
 
Since the daughter particles constitute massless radiation, we can 
rewrite the expression for the evolution of the daughter distribution in 
Eq.~(\ref{BE}) in terms of the background daughter energy density 
$\bar{\rho}_{D}\equiv 4\pi a^{-4} E_{D}^0$ and the background mother 
number density $\bar{n}_{M}\equiv 4\pi a^{-3} N_{M}^0 $, where $E^0_{D}$ 
and $N^0_M$ are defined as in Eq.~(\ref{EDNM}) after expanding out 
$f_{M}$ and $f_{D_i}$ as in Eq.~(\ref{fperturb}),
 \begin{equation}
\label{IBD} 
\frac{\partial \bar{\rho}_{D}}{\partial\tau} + 4 a H \bar{\rho}_{D} = a \Gamma M \bar{n}_{M} .
 \end{equation}
 The right-hand side of the Eq.~(\ref{IBD}) is exactly the same as in 
the case of cold dark matter decay. While mother particles that have 
higher momentum have more energy, they also decay more slowly due to 
time-dilation in the cosmic frame. This perfect cancellation between 
relativistic energy and time-dilation is neatly encapsulated in the 
simplification $\epsilon_M/\gamma=Ma$ that was used in obtaining 
Eq.~(\ref{EM}).

\subsection{Perturbations: First Order}

In the synchronous gauge, the metric perturbations can be parametrized as, 
 \bea
ds^2=a(\tau)^2\left[-d\tau^2+\left(\delta_{ij}+H_{ij}\right)dx^idx^j\right],
 \eea
 where $d\tau=dt/a(\tau)$ and the indices $i$ and $j$ run over the three 
spatial coordinates, $(i,j=1,2,3)$. It is convenient to work in Fourier 
space,
 \bea
H_{ij}(\vec k,\tau)=\hat k_i\hat k_j h(\vec k,\tau)+\left(\hat k_i\hat k_j -\frac{1}{3}\delta_{ij}\right)6\eta(\vec k,\tau) \;,
 \eea
 where $\vec k$ is conjugate to $\vec x$ and $\hat k$ is the unit 
vector. In Fourier space the first order terms in Eq.~(\ref{BE}) for the mother particle can be 
collected as,
 \begin{eqnarray}\label{FO}
\Delta f_M' + i \frac{qk}{\epsilon_M} P_1(\mu)\Delta f_M + q\frac{\partial f_M^0}{\partial q}[-\frac{h'}{6}-\frac{P_2(\mu)}{3}(h' + 6\eta')]&=&-a^2\frac{\Gamma M}{\epsilon_M}\Delta f_M,
 \end{eqnarray}
where $\mu\equiv \hat{k}\cdot\hat{n}$ and $ P_l(\mu)$ are the Legendre 
polynomials.

As usual, we can expand the angular dependence of the 
perturbations as a series in Legendre polynomials,
 \begin{equation}
   X(..., \vec{k},\hat{n})=\sum_{l=0}^{\infty}(-i)^l (2l + 1) X_l(..., k)P_l(\hat{k}\cdot\hat{n}).
 \end{equation}
 Here $X$ represents any of the perturbations $\Delta f_{M,D_j}$, 
$\Delta E_{D}$ or $\Delta N_M$, which are defined as in 
Eqs.~(\ref{fperturb}) and (\ref{EDNM}). Exploiting the orthonormality of 
the Legendre polynomials, we arrive at a Boltzmann hierarchy of moments 
in which any moment is related only to its neighboring moments. The 
diminishing importance of the higher moments allows us to cutoff the 
calculation at some $l = l_{max}$, where the choice of $l_{max}$ depends 
on our desired level of accuracy. We use the improved truncation scheme 
from Ref.~\cite{Ma:1995ey}, which has been generalized to spatial 
curvature in Ref.~\cite{Lesgourgues13}.

The Boltzmann hierarchy for the perturbations of the mother particle becomes,
 \begin{eqnarray}\label{MPert}
\Delta f_{M(0)}' &=& -\frac{qk}{\epsilon_M}\Delta f_{M(1)} + \frac{h'}{6} q\frac{\partial f^0_M}{\partial q} - \frac{a^2 \Gamma M}{\epsilon_M}\Delta f_{M(0)},\nonumber\\
\Delta f_{M(1)}'&=&\frac{qk}{3\epsilon_M}(\Delta f_{M(0)}-2\Delta f_{M(2)}) - \frac{a^2 \Gamma M}{\epsilon_M}\Delta f_{M(1)},\nonumber\\
\Delta f_{M(2)}'&=& \frac{qk}{5\epsilon_M}(2\Delta f_{M(1)}-3 \Delta f_{M(3)})-(\frac{1}{15}h' + \frac{2}{5}\eta')q\frac{\partial f^0_M}{\partial q}- \frac{a^2 \Gamma M}{\epsilon_M}\Delta f_{M(2)},\nonumber\\
\Delta f_{M(l)}'&=& \frac{qk}{(2l+1)\epsilon_M}[l\Delta f_{M(l-1)}-(l+1)\Delta f_{M(l+1)}] - \frac{a^2 \Gamma M}{\epsilon_M}\Delta f_{M(l)},~~l\ge 3.
 \end{eqnarray}  
In the limit that the decay term is set to 
zero, these equations reduce to the standard equations for massive 
neutrinos in the synchronous gauge~\cite{Ma:1995ey}, as expected. 

For the Boltzmann hierarchy of daughter particles, we integrate with 
respect to $\int \dbar^3\vec{q}_{D_j}q_{D_j}P_l(\mu_{D_j})$ on both 
sides of Eq.~(\ref{BE}) for each daughter particle and add them up. The 
collision term becomes
 \begin{align}
\sum_{l^{'}}(-i)^{l^{'}} (2l^{'} + 1)a^2 \int \frac{\dbar^3 \vec q_M}{2\epsilon_M}\prod_i\frac{\dbar^3\vec q_{Di}}{2\epsilon_{Di}} (\sum_j \epsilon_{Dj})(2\pi)^4 \delta^{(4)}(\vec q_M-\Sigma_i \vec q_{Di}) \Delta f_{M(l^{'})}P_l(\mu_{D_j})P_{l^{'}}(\mu_M).\nonumber
 \end{align}
 Again, our focus is on the case in which the mother particle decays 
after becoming non-relativistic. Then, up to corrections of order 
${q_M}/{(Ma)}$ arising from the motion of the mother particle, the decay 
into daughters is isotropic, so that there is no correlation between the 
directions of the mother and daughter momenta ($\hat{n}_{M,D}$). Given 
that the perturbations of the daughter particles give only a small 
contribution to structure formation, we can ignore this subleading 
correction in ${q_M}/{(Ma)}$ and assume that $\mu_M$ and $\mu_{D_j}$ are 
uncorrelated. In this case, the angular integrals over 
the Legendre polynomials can be performed independently, so that
 \barray
\sum_{l^{'}}(-i)^{l^{'}} (2l^{'} + 1)\int_{-1}^1 d\mu_{D_j}P_l(\mu_{D_j})\int_{-1}^1 d\mu_MP_{l^{'}}(\mu_M)=
0\quad ({\rm if}\,\,l\,\,{\rm or}\,\,l'>0)\,.
 \earray
 This implies that in the daughter equations, only the zeroth moment of 
the source term from the mother particle decay ($\Delta f_{M(0)}$) 
survives in the limit of non-relativistic decay. The source term shows 
up in the equation for $\Delta f'_{D(0)}$. We can therefore take 
$l=l^{'}=0$ and simplify the collision term to get a source term similar 
to that in Eq.~(\ref{EM}), but with $f_{M}^0$ replaced by the 
perturbation $\Delta f_{M(0)}$. Therefore, the Boltzmann hierarchy for 
the daughter energy perturbations, $\Delta E_{D(l)}$, in terms of the 
$\Delta N_{M(l)}$ and the metric perturbations $h$ and $\eta$ is given 
by,
 \begin{eqnarray}\label{DPert}
\Delta E'_{{D}(0)}&=&-k \Delta E_{{D}(1)} - \frac{2}{3} h' E_{D}^0 + a^2 M \Gamma \Delta N_{M(0)},\nonumber\\
\Delta E'_{{D}(1)}&=&\frac{k}{3}\Delta E_{{D}(0)}-\frac{2k}{3} \Delta E_{{D}(2)},\nonumber\\
\Delta E'_{{D}(2)}&=&\frac{2k}{5} \Delta E_{{D}(1)} -\frac{3k}{5}\Delta E_{{D}(3)} + \frac{4}{15}E^0_{D} (h' + 6 \eta'),\nonumber\\
\Delta E'_{{D}(l)}&=&\frac{k}{2l + 1}[l \Delta E_{{D}(l-1)}-(l+1)\Delta E_{{D}(l+1)}],~~~~ l\ge3.
 \end{eqnarray} 
 Similar equations can also be found 
in~\cite{Bharadwaj:1997dz,Wang:2012eka,Audren:2014bca,Poulin:2016nat}. 
Again, we neglect the source terms with $\Delta N_{M(l>0)}$ due to the 
additional $q_M/(Ma)$ suppressions in these terms. Other quantities such as 
the overdensity, perturbed pressure, energy flux/velocity-divergence, 
and shear stress can be calculated from these moments in the usual 
manner, to be fed into the perturbed Einstein field equations as 
detailed in \cite{Ma:1995ey}.

\section{Cosmological Signals of Neutrino Decay}

In this section we determine the impact of decaying neutrinos on the 
matter power spectrum and on CMB lensing. In Sec.~\ref{sec:signature}, 
we solve the Boltzmann equations of the previous section numerically 
using {\sf CLASS}, and determine the matter power spectrum and the CMB 
lensing potential $C^{\phi\phi}_\ell$ as a function of the neutrino mass 
and lifetime. This allows us to establish numerically that there is 
indeed a degeneracy in the matter power spectrum between neutrino mass 
and lifetime. In Sec.~\ref{sec:analytical}, we determine the matter 
power spectrum analytically, after making certain well-motivated 
approximations. We show that the results closely reproduce those based 
on the numerical study, and admit a physical interpretation of the 
effects of decaying neutrinos.

\subsection{Numerical Results}\label{sec:signature}

To simplify the analysis, we assume that the three neutrinos have 
degenerate masses and lifetimes. This extends the parameter space of the 
$\Lambda$CDM model to include two additional parameters; the sum of 
neutrino masses, $\sum m_{\nu}$, and the logarithm of the decay width, 
$\log_{10}\Gamma_\nu$. In our analysis, we fix the following 
cosmological parameters to their central values from the {\it Planck} 
2015 TT, TE, EE+low-P data:$\{\omega_{b} = 0.022032,~\omega_{\rm cdm} = 
0.12038,~ {\rm ln}(10^{10}A_s)= 3.052,~ n_s = 0.96229,~ \tau_{\rm reio} 
= 0.0648 \}$.  The impact of neutrino masses on the matter power 
spectrum looks different depending on whether $\theta_s$ or $H_0$ is 
kept fixed~\cite{Archidiacono:2016lnv}. This is because, to keep 
$\theta_s$ fixed, $H_0$ must be adjusted within {\sc CLASS}, leading to 
an overall shift of the matter power spectrum. While fixing $H_0$ is 
more conventional, fixing $\theta_s$ gives a better reflection of the 
constraining effects of a combined analysis of CMB+LSS data, since CMB 
data pins $\theta_s$ down very precisely. In the following, we will show 
results with either $H_0=67.56$ km/s/Mpc or $100\times\theta_s = 1.043$, 
explicitly stating in each case what convention is chosen.

 Since the galaxy power spectrum is known to trace the CDM and baryon 
overdensities, we focus on the power spectrum
 \beq
P_{cb}(k)=\left\langle\frac{\delta\rho_{cb}}{\bar{\rho}_{cb}}\frac{\delta\rho_{cb}}{\bar{\rho}_{cb}}\right\rangle,
 \eeq
 where $\bar{\rho}_{cb}$ ($\delta\rho_{cb}$) is the average 
(perturbation) of the sum of CDM and baryon energy densities.\footnote{ 
Note that this is different from the matter power spectrum 
conventionally defined as $P_m=\langle 
\left[(\delta\rho_{cb}+\delta\rho_{\nu})/(\bar{\rho}_{cb}+\bar{\rho}_{\nu})\right]^2\rangle$, 
which differs from $P_{cb}$ by an extra factor 
$[\bar{\rho}_{cb}/(\bar{\rho}_{cb}+\bar{\rho}_{\nu})]^2$ .
 }
In Fig.~\ref{SignalHO}, we display the residuals of $P_{cb}$ (left) and 
the CMB lensing potential (right) with respect to the case of massless 
neutrinos for $\sum m_\nu$ fixed at 0.25 eV, keeping the value of $H_0$ 
fixed. We compare three different values of $\Gamma_{\nu}$ and the 
limiting case of stable neutrinos. The curves run from top to bottom in 
order of decreasing $\Gamma_{\nu}$. The analytic results are shown as 
dashed lines in the plot, and are seen to agree reasonably well at large 
$k$ or $\ell$ with the numerical results, shown as solid lines. These 
plots demonstrate that the main effect of a non-zero decay rate of 
neutrinos is to reduce the power suppression at large $k$ arising from 
their mass. Moreover, they establish that the gravitational effects of 
unstable relic neutrinos can indeed give rise to observable signals in 
LSS, provided that the decays occur sufficiently long after the 
neutrinos have become non-relativistic.

\begin{figure}
\begin{center}
\includegraphics[scale=0.355]{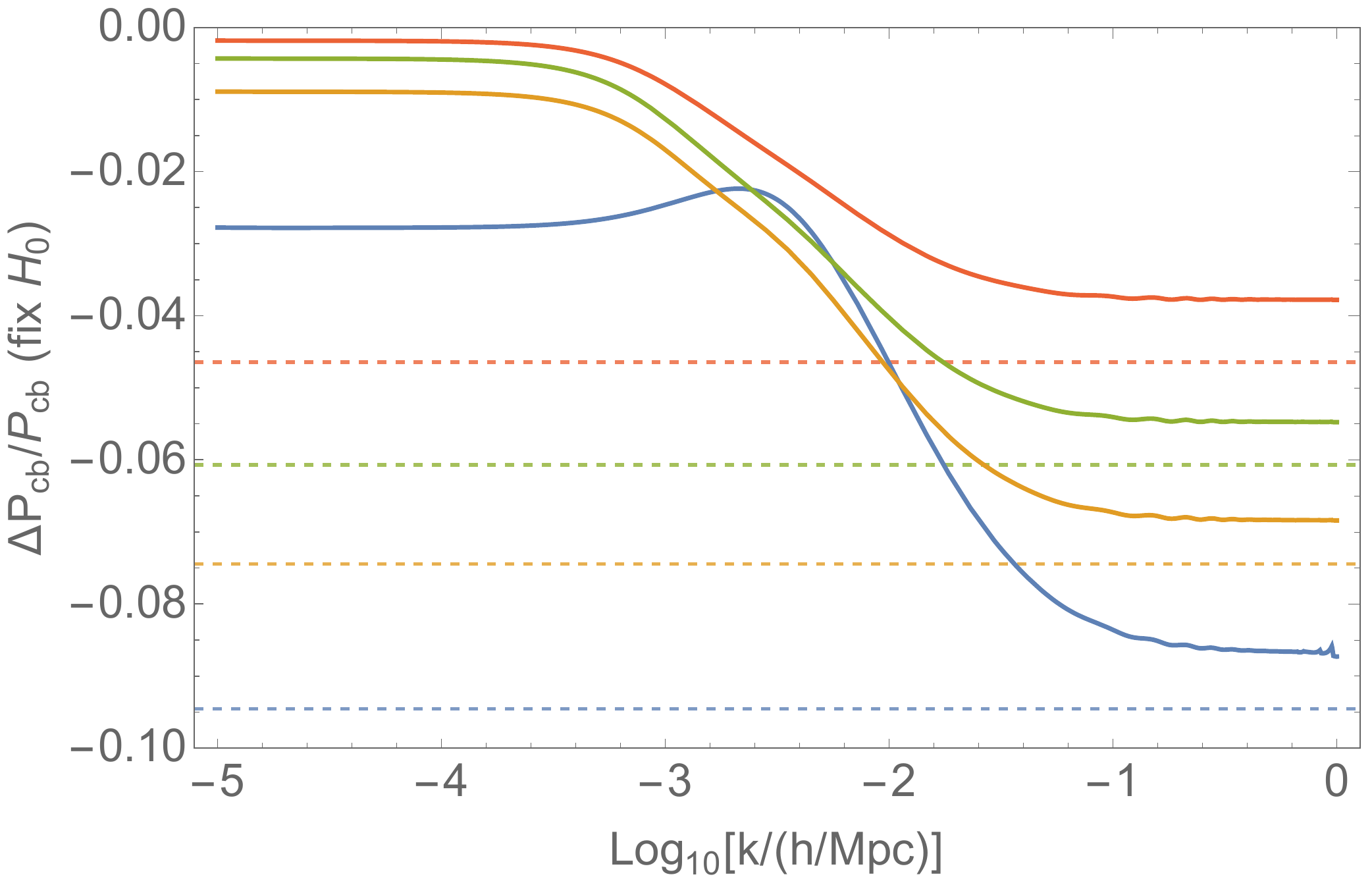}\,\,\,\,
\includegraphics[scale=0.34]{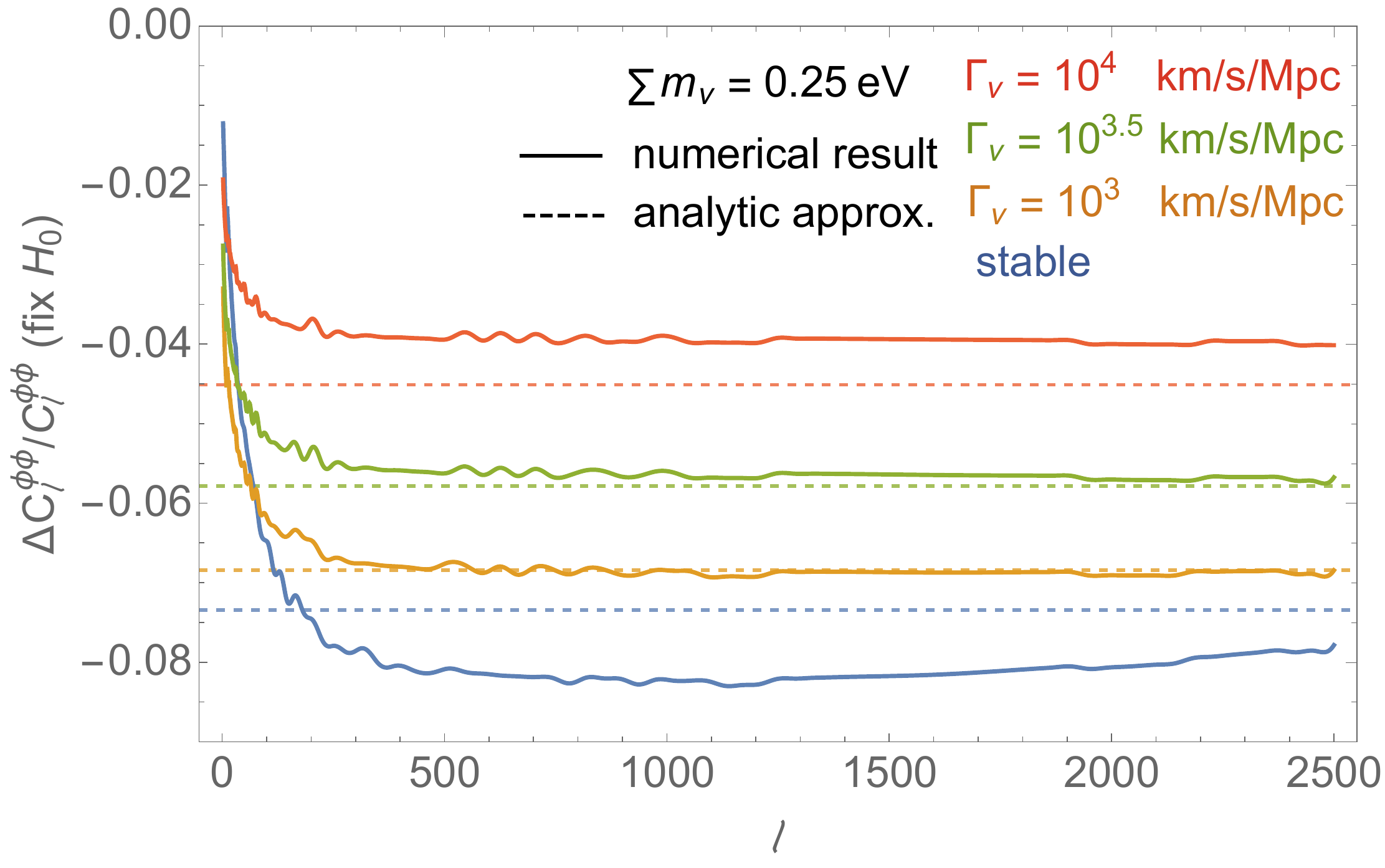}
\end{center}
 \caption{ 
 Plots of the fractional difference in the CDM+Baryon power spectrum 
$P_{cb}$ (\textbf{left}) and CMB-lensing potential $C^{\phi\phi}_\ell$ 
(\textbf{right}) for various decaying (and stable) massive neutrino 
scenarios with respect to the case of massless neutrinos. The solid 
lines show the results from numerical simulations of the decaying 
neutrino scenario for three values of the decay width, $\Gamma_{\nu}$= 
$10^{4.0,~3.5,~3.0}$ (km/s/Mpc) (top to bottom), and also the stable 
neutrino scenario, holding $\sum m_\nu =0.25$ eV and $H_0$ fixed. The 
dashed lines represent the corresponding analytic estimates from 
Sec.~\ref{sec:analytical}.
 }
 \label{SignalHO}
 \end{figure}

\begin{figure}
\hskip-0.2cm
\includegraphics[scale=0.37]{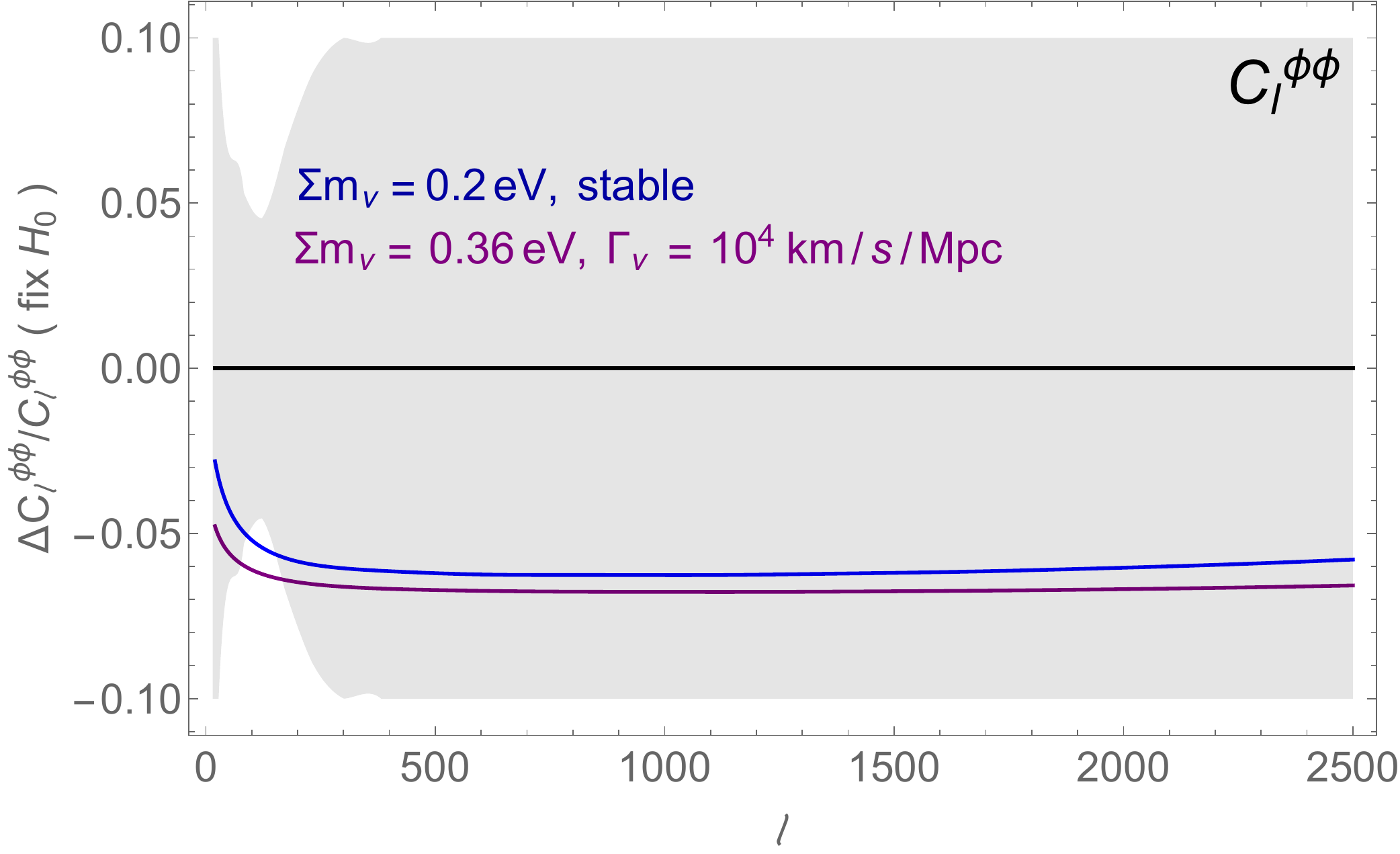}\,\,
\includegraphics[scale=0.362]{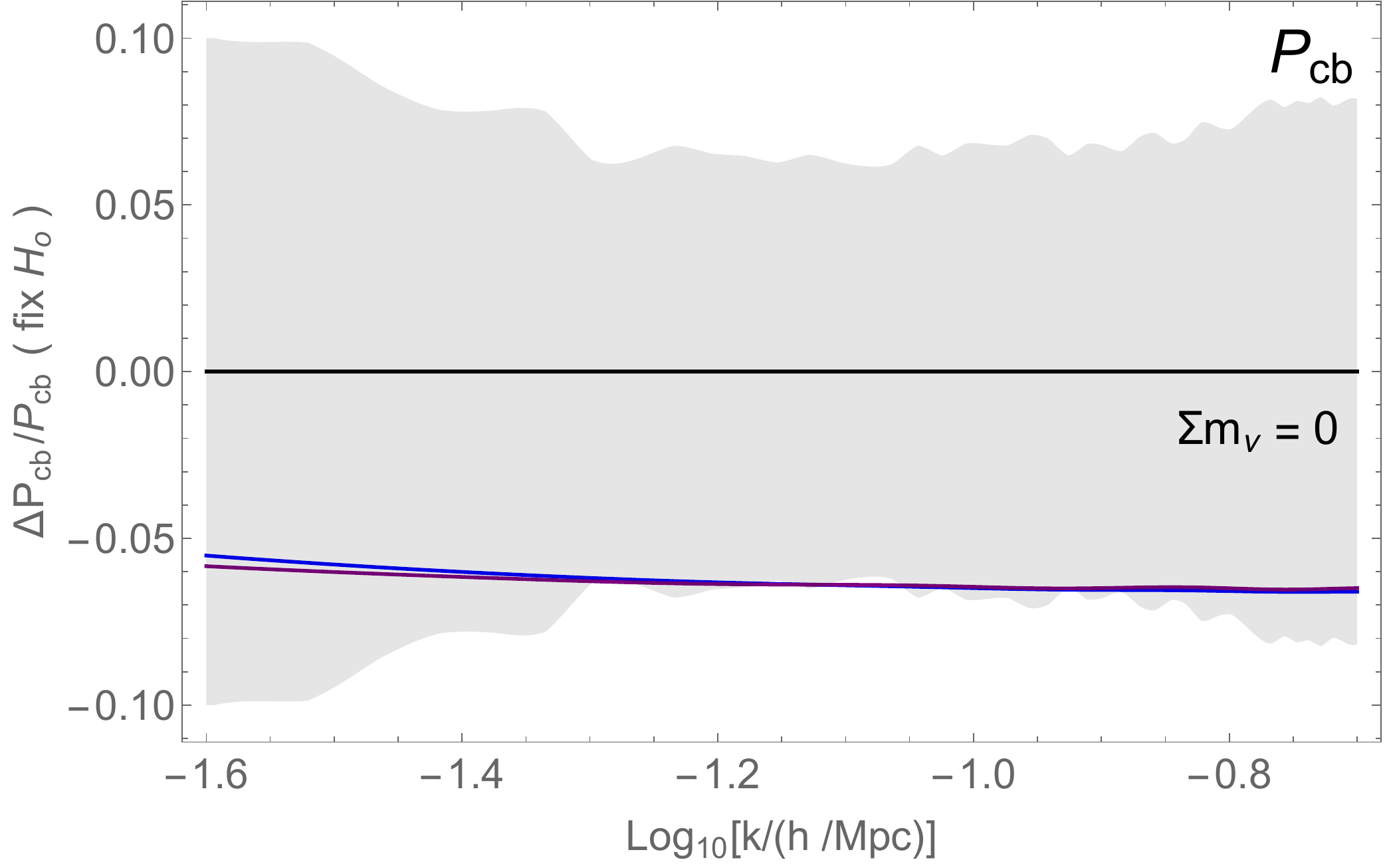}
 \caption{
 The fractional differences in the CMB-lensing potential 
$C^{\phi\phi}_\ell$ (\textbf{left}), CDM+Baryon power spectrum 
$P_{cb}$ (\textbf{right}) for an unstable (purple) and a 
stable (blue) neutrino scenario with respect to the case of massless 
neutrinos (black) at fixed $H_0$. The grey regions show the  $1\sigma$ uncertainties from {\em Planck} and SDSS DR7 respectively.  
}
\label{ratiocompare_fixH0}
\end{figure}

\begin{figure}
\hskip-0.2cm
\includegraphics[scale=0.366]{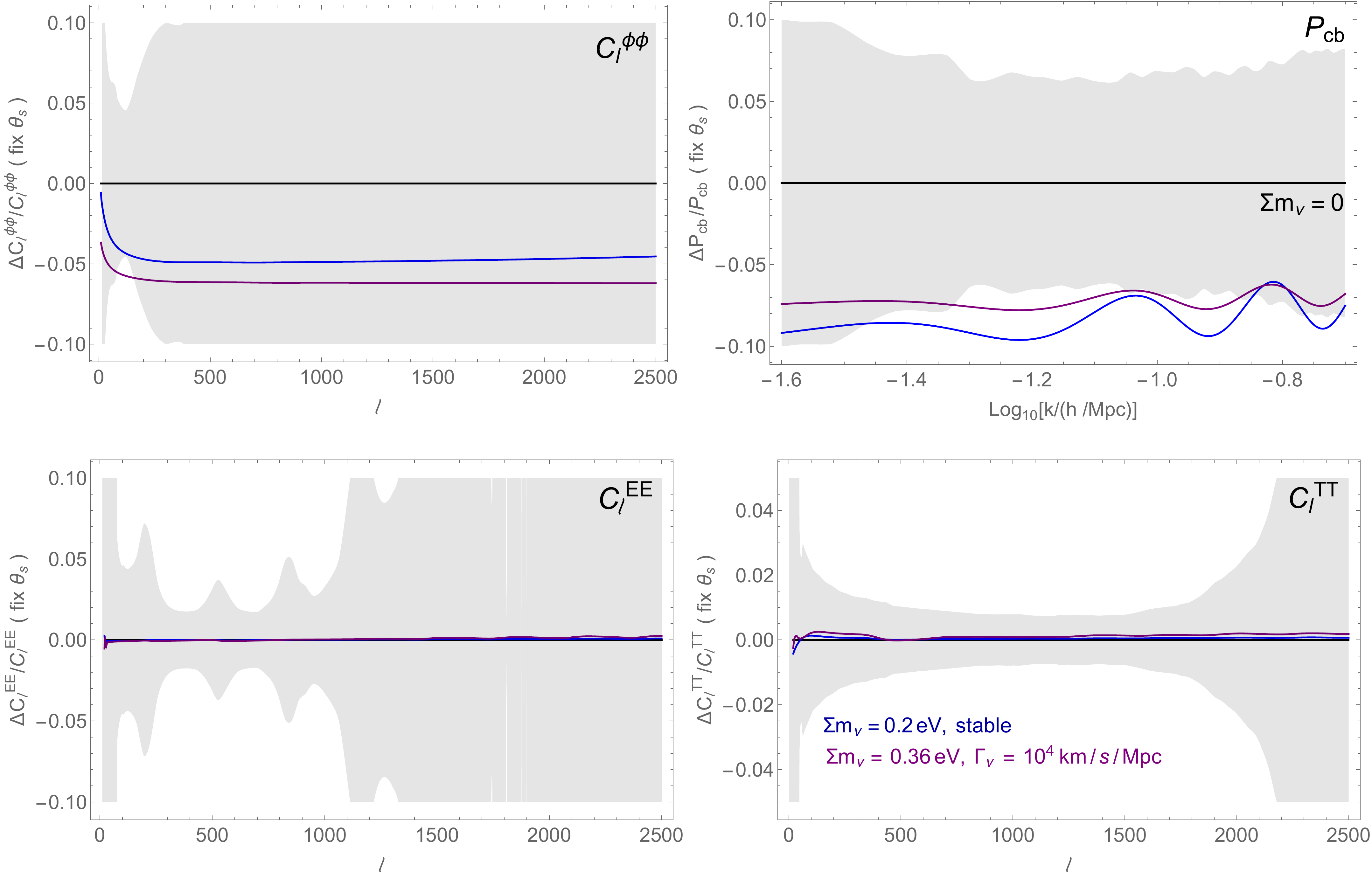}
 \caption{
 The fractional differences in the CMB-lensing potential 
$C^{\phi\phi}_\ell$ (\textbf{top left}), CDM+Baryon power spectrum 
$P_{cb}$ (\textbf{top right}), $C_\ell^{TT}$ (\textbf{bottom left}), and 
$C_\ell^{EE}$ (\textbf{bottom right}) for an unstable (purple) and a 
stable (blue) neutrino scenario with respect to the case of massless 
neutrinos (black) at fixed $\theta_s$. The grey regions show the  $1\sigma$ uncertainties from {\em Planck} and SDSS DR7 respectively. }
\label{ratiocompare_fixtheta}
\end{figure}

Because of the effects of nonlinearities at large $k$ (small scales) and 
cosmic variance at small $k$ (large scales), current experiments are sensitive 
only to a narrow range of $k$ in the neighborhood of $0.1 h/\textrm{Mpc}$. 
We see from Fig.~\ref{SignalHO} that in this region there are no 
qualitative features in $P_{cb}|_{z=0}$ or $C^{\phi\phi}_\ell$ that 
would allow unstable neutrinos to be distinguished from stable ones. 
Although $P_{cb}|_{z=0}$ and $C^{\phi\phi}_\ell$ are more suppressed in 
the stable case, as expected, this effect can be mimicked if the the 
neutrino masses in the unstable scenario are suitably heavier. This 
results in a strong parameter degeneracy between the neutrino lifetime 
and the sum of neutrino masses as determined from $P_{cb}|_{z=0}$ and 
$C^{\phi\phi}_\ell$.

In Fig.~\ref{ratiocompare_fixH0} we show an explicit example of the 
degeneracy between mass and lifetime in the values of $P_{cb}$ and 
$C_{\ell}^{\phi\phi}$ at fixed $H_0$. We consider a model with stable 
neutrinos of mass $\sum m_{\nu}\!=\!0.2$ eV, and a different model with 
unstable neutrinos of mass $\sum m_{\nu}\!=\!0.36$ eV and width 
$\Gamma_{\nu}\!=\!10^{4}$ km/s/Mpc. In the $P_{cb}(z=0)$ case, we see 
from the figure that the blue (stable neutrino) and purple (unstable 
neutrino) curves cannot be distinguished by measurements such as SDSS 
DR7 (used later in sec.~\ref{sec:MCstudy}), whose sensitivity is shown 
in grey. However, we note that the lensing power spectrum can 
potentially help in breaking the degeneracy, because it receives its 
dominant contribution at higher $z\approx 3$~\cite{Bernardeau:1996aa}. 
We will explore the possibility of breaking the degeneracy by using next 
generation measurements at different redshifts in future work.

Finally, we show in Figs.~\ref{ratiocompare_fixtheta} the effects of 
neutrino masses and decay at fixed $\theta_s$ on $P_{cb}$, 
$C_{\ell}^{\phi\phi}$ and $C_{\ell}^{TT,EE}$. This fixes the peak 
locations in the CMB power spectra and only generates negligible 
deviations away from the massless neutrino case in 
$C_{\ell}^{TT,EE}$~\cite{Archidiacono:2016lnv}. The same choices of 
parameters, however, do generate sizeable deviations in 
$C_{\ell}^{\phi\phi}$ and $P_{cb}$ away from the massless neutrino case 
that are close to the current sensitivities. This demonstrates that as 
expected, for sub-eV $\sum m_\nu$, it is the CMB-lensing and matter 
power spectrum measurements that provide the constraining power. 
Additionally, note that the change in $H_0$ required to keep $\theta_s$ 
fixed leads to an overall shift of $P_{cb}$. This makes the BAO in the 
three models out of phase and leads to small oscillations at large $k$ 
on top of the power suppression.

\subsection{Analytic Understanding}\label{sec:analytical}

In this section we provide an analytic derivation of the effects of 
neutrino decay on CMB and LSS observables. We begin by showing how the 
results in the literature for the effects of massive neutrinos on the 
matter power spectrum ($P_{cb}(k)$) and CMB lensing 
($C_{\ell}^{\phi\phi}$) can be reproduced analytically. We improve on 
the existing analytical treatment of the cosmological effects of massive 
neutrinos by taking into account their momentum distribution. We then 
build on this to derive an expression for the evolution of overdensities 
in scenarios with unstable neutrinos.

Once neutrinos become non-relativistic, their contribution to the 
background energy density leads to an increase the Hubble rate, leaving 
less time for structure formation as compared to a universe with 
massless neutrinos. The net result is an overall suppression of power at 
small scales in the matter power spectrum. The size of this effect can 
be determined by studying the evolution of density perturbations. 
Consider $\delta_{i}=\delta \rho_{i}/\bar{\rho}_{i}$ for particle 
species $i$, for a mode that is already deep inside the horizon when 
neutrinos become non-relativistic at $z=z_{\rm nr}$. In the matter 
dominated era, the Einstein equation for the density perturbation 
with wavenumber $k$ can be approximated as
 \begin{equation}\label{eq:einstein}
k^2\phi \approx -4\pi\,Ga^2(\delta_{{cb}}\,\bar{\rho}_{{cb}}+\delta_{\nu}\,\bar{\rho}_{\nu}).
\end{equation}
 Here $\phi$ is the metric perturbation in the conformal Newtonian 
gauge~\cite{Ma:1995ey}.\footnote{We use the metric 
$ds^2=a^2(\tau)[-(1+2\psi)d\tau^2+(1-2\phi)\delta_{ij}dx^idx^j]$ and 
approximate $\psi=-\phi$, ignoring the small correction arising from the 
presence of free streaming radiation.} We assume baryons have already 
decoupled from photons. This allows us to combine the baryon 
contribution to the matter density with that of CDM to simplify the 
discussion. Since $\delta_{\nu}\ll\delta_{cb}$ for perturbation 
modes that enter the horizon before $z_{\rm nr}$, we can write,
 \begin{equation}\label{eq:phiapprox}
k^2\phi \approx -\frac{6}{\tau}\,\left(1-\frac{\bar{\rho}_{\nu}(\tau)}{\bar{\rho}_{\rm tot}(\tau)}\right)\,\delta_{cb},
 \end{equation}
 where $\tau$ is the comoving time and $\bar{\rho}_{\rm tot}\equiv 
\bar{\rho}_{cb}+\bar{\rho}_{\nu}$. Inserting this expression into the 
Boltzmann equation for CDM perturbations yields,
 \begin{equation}\label{eq:eq}
\ddot{\delta}_{cb}+\frac{2}{\tau}\dot{\delta}_{cb}-\frac{6}{\tau^2}\left(1-f_{\nu}(\tau)\right)\delta_{cb}=0,
\qquad f_{\nu}(\tau)=\frac{\bar{\rho}_{\nu}(\tau)}{\bar{\rho}_{\rm tot}(\tau)}.
 \end{equation}
 where the dots represent derivatives with respect to $\tau$. Deep in 
the matter dominated era, neutrinos only contribute up to a few percent 
of the total energy density. Therefore, throughout this derivation, we 
work to leading order in $f_{\nu}\,\,(\ll1)$. We look for a solution of 
the form,
 \begin{equation}
\delta_{cb}=\delta_{{cb},i} h(\tau)
\left(\frac{\tau}{\tau_i}\right)^2\exp\left[-\frac{6}{5}\displaystyle{\int_{\tau_i}^{\tau}}\frac{d\hat{\tau}}{\hat{\tau}}f_{\nu}(\hat{\tau})\right]
 \end{equation}
 where now the function $h(\tau)$ is to be determined.
Inserting this expression into Eq.~(\ref{eq:eq}) and dropping the term proportional to $f_{\nu}^2$, we obtain the following
differential equation for $h(\tau)$.
 \begin{equation}\label{eq:eq2}
\tau\,\ddot{h}+6\dot{h}-\frac{6}{5}h\,\dot{f}=0\,\,.
 \end{equation}
 Thus far we have not made any assumption about the redshift dependence of $f_{\nu}$. For massless or ultrarelativistic neutrinos in the matter dominated era, we have
 \begin{equation}
f_{\nu}(\tau)=f_{\nu,i}\left(\frac{\tau_i}{\tau}\right)^2,
 \end{equation}
 In this case we can solve for the function $h(\tau)$ as,
 \begin{equation}\label{eq:hfunc}
h^{\not{m}_{\nu}}(\tau)=\exp\left[k\,\displaystyle{\int_{\tau_i}^{\tau}}\,d\hat{\tau}\,\dot{f}^{\not{m}_{\nu}}_{\nu}(\hat{\tau})\right]
                       =\exp\left[\frac{2}{5}\left(f^{\not{m}_{\nu}}_{\nu}(\tau_i)-f^{\not{m}_{\nu}}_{\nu}(\tau)\right)\right].
 \end{equation}
 This leads to the following approximate solution for perturbations in the case of massless or ultrarelativistic neutrinos,
 \begin{eqnarray}
\delta^{\not{m}_{\nu}}_{cb}(\tau)=\delta_{{cb},i}\left(\frac{\tau}{\tau_i}\right)^2
                                       \exp\left[-\frac{6}{5}\displaystyle{\int_{\tau_i}^{\tau}}\frac{d\hat{\tau}}{\hat{\tau}}f^{\not{m}_{\nu}}_{\nu}(\hat{\tau})\right]\,
                                       h^{\not{m}_{\nu}}(\tau).
 \end{eqnarray}
 In the limit that neutrinos are non-relativistic, $f_{\nu}(\tau)$ goes to a constant value. Then Eq.~(\ref{eq:eq2}) admits a solution where $h(\tau)$ is constant.  
 This implies that in the case of massive neutrinos, the $h$-function can be approximated as
 \begin{equation}
h^{m_{\nu}}(\tau)=\exp\left\{\frac{2}{5}\left[f^{\not{m}_{\nu}}_{\nu}(\tau_i)-f^{\not{m}_{\nu}}_{\nu}\left({\rm min}(\tau,\tau_{\rm nr})\right)\right]\right\}.
 \end{equation}
 The result is almost identical to Eq.~(\ref{eq:hfunc}) since in both cases the exponent is dominated by $f^{\not{m}_{\nu}}_{\nu}(\tau_i)$ (and $f^{\not{m}_{\nu}}_{\nu}$ after
$\tau>\tau_{\rm nr}$ is much smaller than the expansion parameter $f^{m_{\nu}}_{\nu}$)
 \begin{equation}
h^{m_{\nu}}(\tau)\approx h^{\not{m}_{\nu}}(\tau).
 \end{equation}
 This means that the solution in the case of massive neutrinos can be approximated as,
 \begin{eqnarray}
\delta^{m_{\nu}}_{cb}(\tau)=\delta_{{cb},i}\left(\frac{\tau}{\tau_i}\right)^2
                                 \exp\left[-\frac{6}{5}\displaystyle{\int_{\tau_i}^{\tau}}\frac{d\hat{\tau}}{\hat{\tau}}f^{m_{\nu}}_{\nu}(\hat{\tau})\right]\,h^{\not{m}_{\nu}}(\tau).
 \end{eqnarray}
 Then the ratio of the perturbations in the two cases is given by,
 \begin{eqnarray}
 \label{eq:ratio1}
\frac{\delta^{m_{\nu}}_{cb}(\tau)}{\delta^{\not{m}_{\nu}}_{cb}(\tau)}
=\exp\left[-\frac{6}{5}\displaystyle{\int_{\tau_i}^{\tau}}\frac{d\hat{\tau}}{\hat{\tau}}\left(f^{m_{\nu}}_{\nu}(\hat{\tau})-f^{\not{m}_{\nu}}_{\nu}(\hat{\tau})\right)\right].
 \end{eqnarray}
 This ratio can be expressed in terms of the scale factor as,
 \begin{equation}\label{eq:deltaratio}
\frac{\delta^{m_{\nu}}_{cb}(a)}{\delta^{\not{m_{\nu}}}_{cb}(a)}\approx\frac{\delta^{m_{\nu}}_{cb}(a_i)}{\delta^{\not{m_{\nu}}}_{cb}(a_i)}\exp\left[-\frac{3}{5}\displaystyle{\int_{a_i}^a}\frac{d a}{a}\frac{\hat{\rho}_{\nu}(a)}{\bar{\rho}_{\rm tot}(a)}\right],
 \end{equation}
 where $\hat{\rho}_{\nu}(a)\equiv\bar{\rho}_{\nu,m_{\nu}}(a)-\bar{\rho}_{\nu,\not{m_{\nu}}}(a)$ represents the difference in the neutrino energy between the two scenarios. 
If all the neutrinos 
are stable and become non-relativistic instantly at $a_i$, $\hat{\rho}_{\nu}(a)/\bar{\rho}_{\rm tot}(a)=\bar{\rho}_{\nu,m_{\nu}}/\bar{\rho}_{\rm tot}$ is a constant, and 
Eq.~(\ref{eq:deltaratio}) recovers the well-known result for the ratio of perturbations in the massive and massless neutrino scenarios,
 \begin{equation}\label{eq:oldestimate}
\frac{\delta^{\,m_{\nu}}_{cb}(a)}{\delta^{\,\not{m_{\nu}}}_{cb}(a)}\propto\left(\frac{a}{a_{i}}\right)^{-\frac{3}{5}\frac{\bar{\rho}_{\nu,m_{\nu}}}{\bar{\rho}_{\rm tot}}}.
 \end{equation}

We can improve on this estimate by incorporating a more precise 
expression for the neutrino energy in Eq.~(\ref{eq:deltaratio}), 
 \begin{equation}
\hat{\rho}_{\nu}(a)=4\pi a^{-4}\displaystyle{\int_{0}^{\infty}}dq\,q^2\left(\sqrt{q^2+m_{\nu}^2a^2}-q\right)\,f(q)\,\,.
 \end{equation}
 Here $q=a\,p_{\nu}$ denotes the neutrino's conformal momentum, and 
$f(q)=\left[e^{q/T_{\nu 0}}+1\right]^{-1}$ represents the momentum 
distribution of neutrinos. $\hat{\rho}_{\nu}(a)$ exhibits non-trivial 
redshift dependence since the neutrino energy goes from being 
radiation-like to being matter-like. In fig.~\ref{fig:PSevolution}, we 
show the evolution of the ratio in Eq.~(\ref{eq:deltaratio}) as a 
function of redshift (black dashed curves) for two different values of 
the neutrino mass. We start our approximation from $a_i=2\times10^{-3}$ 
to make sure we are deep inside the matter dominated era so that the 
assumptions leading to Eq.~(\ref{eq:deltaratio}) are justified.  We 
stress, however, that the result is quite insensitive to order one 
changes in $a_i$. As we can see, Eq.~(\ref{eq:deltaratio}) is a good approximation to the full numerical results (black solid 
curves), and describes the evolution of the $\delta_{cb}$ ratio from the 
relativistic to the non-relativistic regime much better than the 
approximation based on Eq.~(\ref{eq:oldestimate}) (black dotted curves). 
Using this, we can estimate the ratio of the power spectrum between the 
two scenarios,
 \begin{equation}\label{eq:PratioWDM}
\frac{P_{{ cb},m_{\nu}}}{P_{{ cb},\not{m_{\nu}}}}\approx \left(\frac{\delta^{\,m_{\nu}}_{cb}(a_f)}{\delta^{\,\not{m_{\nu}}}_{cb}(a_f)}\right)^2.
 \end{equation}
 The density perturbation grows much slower in the cosmological constant 
dominant era, and we take the final scale factor to be at $a_f=0.7$ for 
a good approximation to the power spectrum ratio today\footnote{We can 
also use $a_f=a\,g(a)$ with the growth function $g(a)$ for a reasonable 
approximation~\cite{Lesgourgues:2018ncw}.}.

\begin{figure}[t]
\begin{center}
\includegraphics[width=10cm]{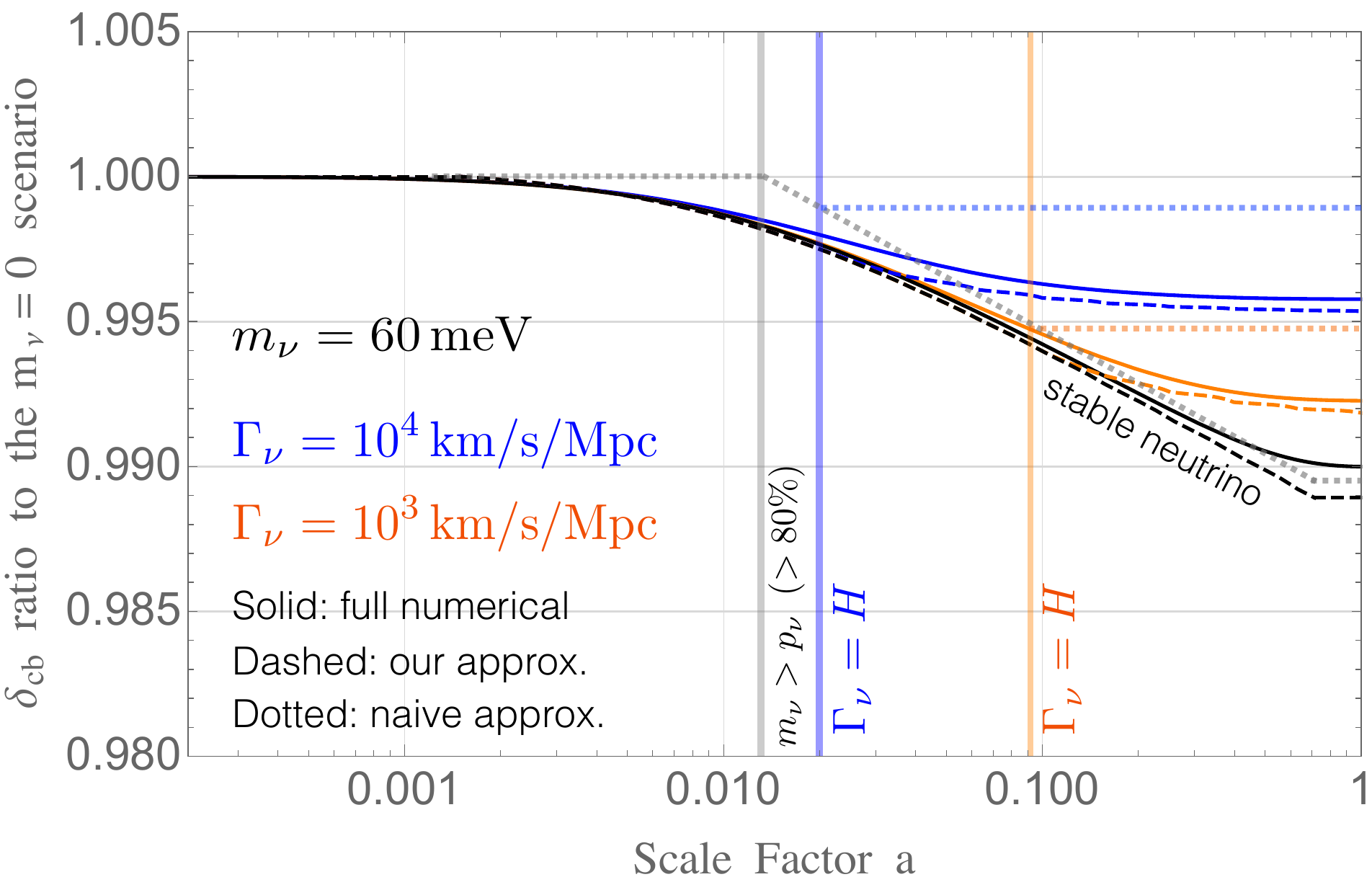}
\\\vspace{1em}
\includegraphics[width=10cm]{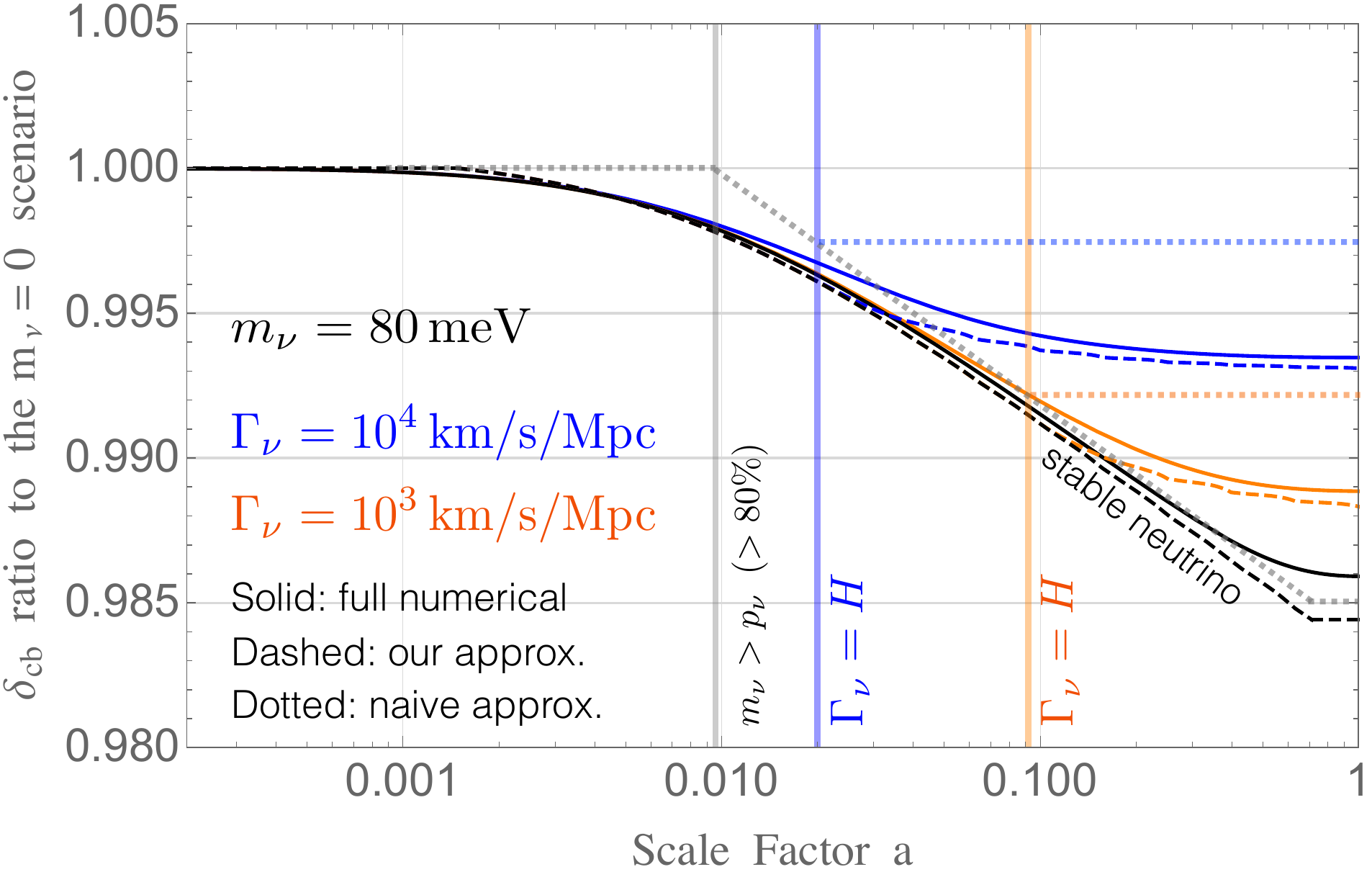}
\end{center}
\caption{
 Evolution of the ratio of the CDM+baryon density perturbation with 
respect to the case of a massless neutrino, 
$\delta_{cb}^{m_{\nu}}/\delta_{cb}^{\not{m_{\nu}}}$. The results are 
shown for the case of a single massive neutrino with $m_{\nu}=60$ meV. 
All the solid curves are obtained from numerical calculations using the 
modified {\sf CLASS} code discussed in Sec.~\ref{sec:numerics}. The 
black curve is for the stable neutrino scenario, and the blue (orange) 
curve is for the neutrino with decay rate $\Gamma_{\nu}=10^4\,(10^3)$ 
km\,/s/Mpc. The dashed curves represent the first approximations to the 
solid curves, based on the derivation in Eq.~(\ref{eq:deltaratio}). The 
dotted curves are based on the approximation method in 
Eq.~(\ref{eq:oldestimate}), where we assume $a_i$ to be the value when 
$80\%$ of neutrinos have their momenta lower than $m_{\nu}$ and 
$a_f=a_{\rm dec}$. As we see, Eq.~(\ref{eq:deltaratio}) provides a much 
better approximation to the full numerical result.}
 \label{fig:PSevolution}
\end{figure}

 We now turn our attention to the effects of massive neutrinos on CMB 
lensing. The difference in the density perturbation $\delta\rho_{cb}$ between the massive and massless neutrino scenarios results in a 
change in the gravity perturbation $\phi$. The photons are therefore 
deflected differently in the CMB lensing process. The correlation 
function of the lensing potential, 
$C_{\ell}^{\phi\phi}\sim\langle\phi\phi\rangle$, parameterizes the size 
of angular deflection of CMB photons. The ratio of
$C_{\ell}^{\phi\phi}$ in the massive neutrino case to that in the 
massless case can be approximated using Limber's 
formula~\cite{1953ApJ...117..134L,Pan:2014xua}
 \begin{equation}
\frac{C_{\ell,\,m_{\nu}}^{\phi\phi}}{C_{\ell,\not{m_{\nu}}}^{\phi\phi}}\approx \frac{\chi_{*}^{m_{\nu}}\displaystyle{\int^{1}_0}{\rm d}x\,\phi^2_{m_{\nu}}\left(\frac{\ell}{x^{m_{\nu}}\chi_*}\right)\,(1-x)^2}{\chi_{*}^{\not{m_{\nu}}}\displaystyle{\int^{1}_0}{\rm d}x\,\phi^2_{\not{m_{\nu}}}\left(\frac{\ell}{x^{\not{m_{\nu}}}\chi_*}\right)\,(1-x)^2}\,\,,\quad x\equiv\frac{\tau_f-\tau}{\tau_f-\tau_*}\,,\quad\chi_*\equiv \tau_f-\tau_*\,\,.
 \end{equation}
 Here $\tau_*\approx 2.8 \times 10^2$ Mpc is the conformal time at last 
scattering, while $\tau_f\approx 1.4\times 10^4$ Mpc is the conformal 
time today. The value of $\tau_f$ differs a bit between the massive and 
massless neutrino scenarios, since the contribution of neutrinos to the 
total energy density is different in the two cases. However, since the 
neutrino mass only results in a significant difference in the 
contributions to the background energy in the short period of time 
between the neutrinos becoming non-relativistic and the universe 
becoming dominated by the cosmological constant, the difference in 
$\chi_*$ between the two scenarios can be neglected. Then, the 
difference between $C_{\ell}^{\phi\phi}$ in the two cases primarily 
arises from differences in the evolution of $\phi$.

 According to the Einstein Eq.~(\ref{eq:einstein}), the ratio of $\phi$ 
between the two scenarios for large $\ell$ modes at a given value of the 
scale factor is,
 \begin{equation}
\frac{\phi_{m_{\nu}}(a)}{\phi_{\not{m_{\nu}}}(a)}\approx\frac{\delta_{cb}^{m_{\nu}}(a)}{\delta_{cb}^{\not{m_{\nu}}}(a)}\,\,.
 \end{equation}
 Since $C_{\ell}^{\phi\phi}$ receives its dominant contribution close to 
$z\approx 3$~\cite{Bernardeau:1996aa}, we can estimate the ratio of the 
$C_{\ell}^{\phi\phi}$ as,
 \begin{equation}\label{eq:CLphiratio}
\frac{C_{\ell,\,m_{\nu}}^{\phi\phi}}{C_{\ell,\not{m_{\nu}}}^{\phi\phi}}\approx \left(\frac{\delta_{cb}^{m_{\nu}}}{\delta_{cb}^{\not{m_{\nu}}}}\right)^2\Bigg|_{z=3}\,\,.
 \end{equation}

Based on a very similar analysis, we can predict the suppression of 
$P_{cb}(k)$ and $C_{\ell}^{\phi\phi}$ for large $k$ and $\ell$ in the 
unstable neutrino case. We consider a scenario with a single massive 
neutrino species that becomes non-relativistic after last scattering and 
decays into dark radiation. After the decay, the energy density of the 
\emph{daughter} particles redshifts more quickly than that of a stable 
neutrino of the same mass as the mother. We work in the instantaneous 
decay approximation and assume that all neutrinos decay at the same 
time, corresponding to a scale factor $a_{\rm dec}$, which is implicitly 
determined by the equation,
 \begin{equation}
\Gamma_{\nu} =H(a_{\rm dec}).
 \end{equation}
 The difference in energy density $\hat{\rho}_{\nu}$ between an unstable 
neutrino and a massless neutrino evolves in a more complicated way than 
in the case of a stable neutrino. The instantaneous decay approximation 
allows us to separate the evolution into two parts. On timescales 
shorter than the proper lifetime of the neutrino, the difference in 
energy density follows the equation,
 \begin{equation}\label{eq:Dnurhoratio1}
\hat{\rho}_{\nu}(a)=4\pi a^{-4}\displaystyle{\int_{0}^{\infty}}dq\,q^2\left(\sqrt{q^2+m_{\nu}^2a^2}-q\right)\,f(q)\,,\quad a<a_{\rm dec}\,.
 \end{equation}
 In the instantaneous decay approximation, the energy density in 
non-relativistic neutrinos is immediately transferred into radiation 
energy at $a_{\rm dec}$. It subsequently redshifts with an extra 
$(a_{\rm dec}/a)$ factor as compared to a non-relativistic neutrino, so that
 \begin{equation}\label{eq:Dnurhoratio2}
\hat{\rho}_{\nu}(a)=4\pi a^{-4}\displaystyle{\int_{0}^{\infty}}dq\,q^2\left[\sqrt{q^2+m_{\nu}^2a^2}\left(\frac{a_{\rm dec}}{a}\right)-q\right]\,f(q)\,,\quad a\geq a_{\rm dec}\,.
\end{equation}
 The ratio of CDM density perturbations in the case of unstable 
neutrinos can be obtained by inserting the energy density ratios in 
Eqs.~(\ref{eq:Dnurhoratio1}) and (\ref{eq:Dnurhoratio2}) into 
Eq.~(\ref{eq:deltaratio}). Then the ratios of $P(k)$ and 
$C^{\phi\phi}_{\ell}$ in the limit of large $k$ and $\ell$ can be 
obtained from Eqs.~(\ref{eq:PratioWDM}) and (\ref{eq:CLphiratio})

In Fig.~\ref{fig:PSevolution}, we show the ratio of $\delta_{cb}$ from 
the numerical calculation described in Sec.~\ref{sec:numerics} for both 
the decaying (blue and orange) and stable (black) neutrinos. The plots 
are for a single massive neutrino with $m_{\nu}=60$ meV (upper) and $80$ 
meV (lower), and a decay rate $\Gamma_{\nu}=10^4\,(10^3)$ km\,/s/Mpc for 
the blue (orange) curves. In this scenario, more than $80\%$ of the 
neutrinos have momenta $p_{\nu}<m_{\nu}$ after $a>0.012$ ($a>0.0096$) 
for $m_{\nu}=60\,(80)$ meV neutrino. It is at this point, when most of 
the neutrinos have become non-relativistic, that the major suppression 
of $\delta_{cb}$ begins. During this period the $\delta_{cb}$ ratio 
drops with the power described in Eq.~(\ref{eq:oldestimate}) (grey 
line). The blue (orange) dotted lines give the value of the 
$\delta_{cb}$-suppression if the later contributions of daughter 
particles to the energy density shown in Eq.~(\ref{eq:Dnurhoratio2}) are 
ignored. As we see, this underestimates the suppression of 
$\delta_{cb}$, showing that the contributions of daughter particles to 
the energy density cannot be neglected. It is clear from the figures 
that Eqs.~(\ref{eq:Dnurhoratio1}) and (\ref{eq:Dnurhoratio2}) provide a 
good description of the $\delta_{cb}$ evolution in unstable neutrino 
scenarios (dashed blue and orange), both before and after neutrino 
decay. This shows that the effects of neutrino decay on the evolution of 
$\delta_{cb}$ on these length scales primarily arise from the 
contributions of the unstable neutrinos and their daughter particles to 
the background energy density, and not from their perturbations.

\section{Current Limits on the Neutrino Mass and Lifetime
from Monte Carlo Analysis}\label{sec:MCstudy}

In this section we perform a Monte Carlo analysis to determine the 
current bounds on the neutrino mass and lifetime. 

\subsection{The Data and Analysis Pipeline}
Our analysis makes use of various combinations of the following datasets.
\begin{itemize}
\item CMB: We include \textit{Planck} 2015 CMB high-$\ell$ TT, TE, and EE and low-$\ell$ TEB power spectra~\cite{Aghanim:2015xee}, as well as  the lensing reconstruction power spectrum ~\cite{Ade:2015zua}.
\item BAO: We use measurements of the volume distance from 6dFGS at $z = 0.106$~\cite{Beutler:2011hx} and the MGS galaxy sample of SDSS at $z = 0.15$~\cite{Ross:2014qpa}.
  We include the anisotropic measurements from the CMASS and LOWZ galaxy samples from the BOSS DR12 at $z = 0.38$, $0.51$, and $0.61$~\cite{Alam:2016hwk}.
\item Growth Function: The BOSS DR12 measurements also include measurements of the growth function $f$, defined by
  \begin{equation}
    f\sigma_8 \equiv \frac{\left[\sigma_8^{(vd)}(z)\right]^2}
                          {\sigma_8^{(dd)}(z)} \ ,
  \end{equation}
  where $\sigma_8^{(vd)}$ measures the smoothed density-velocity correlation, analogous to $\sigma_8 \equiv \sigma_8^{(dd)}$ that measures the smoothed density-density correlation.
\item Pantheon: we use the Pantheon supernovae dataset \cite{Scolnic:2017caz}, which includes measurements of the luminosity distance of 1048 SNe Ia in the redshift range $0.01 < z < 2.3$.  
\item LSS: We use the measurement of the halo power spectrum from the Luminous Red Galaxies SDSS-DR7~\cite{Reid:2009xm}\footnote{More recent measurements are not yet available in {\sc MontePython-v3}. These could naturally make the bounds presented here slightly stronger.} and the tomographic weak lensing power spectrum by KiDS \cite{Kohlinger:2017sxk}.
\end{itemize}
 Our baseline analysis makes use of Planck+BAO+Growth Function+Pantheon 
data (i.e. data that relies on background cosmology or perturbations in 
the linear regime mostly). We then add LSS 
information to gauge the constraining power of such surveys.

Using the public code {\sc 
MontePython-v3}\footnote{https://github.com/brinckmann/montepython\_public} 
\cite{Audren:2012wb,Brinckmann:2018cvx}, we run Monte Carlo Markov chain 
analyses using the Metropolis-Hastings algorithm assuming flat priors 
on all parameters. Our $\Lambda$CDM parameters are,
 \begin{equation*}
  \{\omega_\mathrm{cdm},\omega_b,\theta_s,{\rm ln}(10^{10} A_s),n_s,\tau_\mathrm{reio}\} \,,
\end{equation*}
 to which we add the sum of neutrino masses $\sum m_\nu$ and the logarithm 
of the neutrino lifetime ${\rm Log}_{10}\Gamma_\nu$. In our analysis we 
assume 3 degenerate, unstable neutrino species that decay into dark 
radiation. Although not detailed for brevity, there are many nuisance 
parameters that we analyze together with these cosmological parameters. 
To that end, we employ a Cholesky decomposition to handle the large number 
of nuisance parameters~\cite{Lewis:2013hha}, and use the default priors 
that are provided by {\sc MontePython-v3}. 

\subsection{Current Limits on the Neutrino Mass and Lifetime}

In order to perform meaningful comparisons and to check the accuracy of 
our modified version of {\sf CLASS}, we begin by running the case of 
{\em stable} neutrinos. Our baseline constraint on the neutrino mass, 
obtained with Planck+BAO+Growth Function+Pantheon, is $\sum m_\nu < 0.28$ eV (95\% C.L.). This 
is in good agreement with the result reported in~\cite{Ade:2015xua}. 
The inclusion of SDSS DR7 and KiDS improves the constraint by 
$\sim10\%$, bringing the limit down to $\sum m_\nu < 0.25$ eV (95\% C.L.). 
This constraint when LSS data is included is also 
in good agreement with what is reported in Ref.~\cite{Vagnozzi:2017ovm}.

\begin{figure}
    \centering
    \includegraphics[scale=0.47]{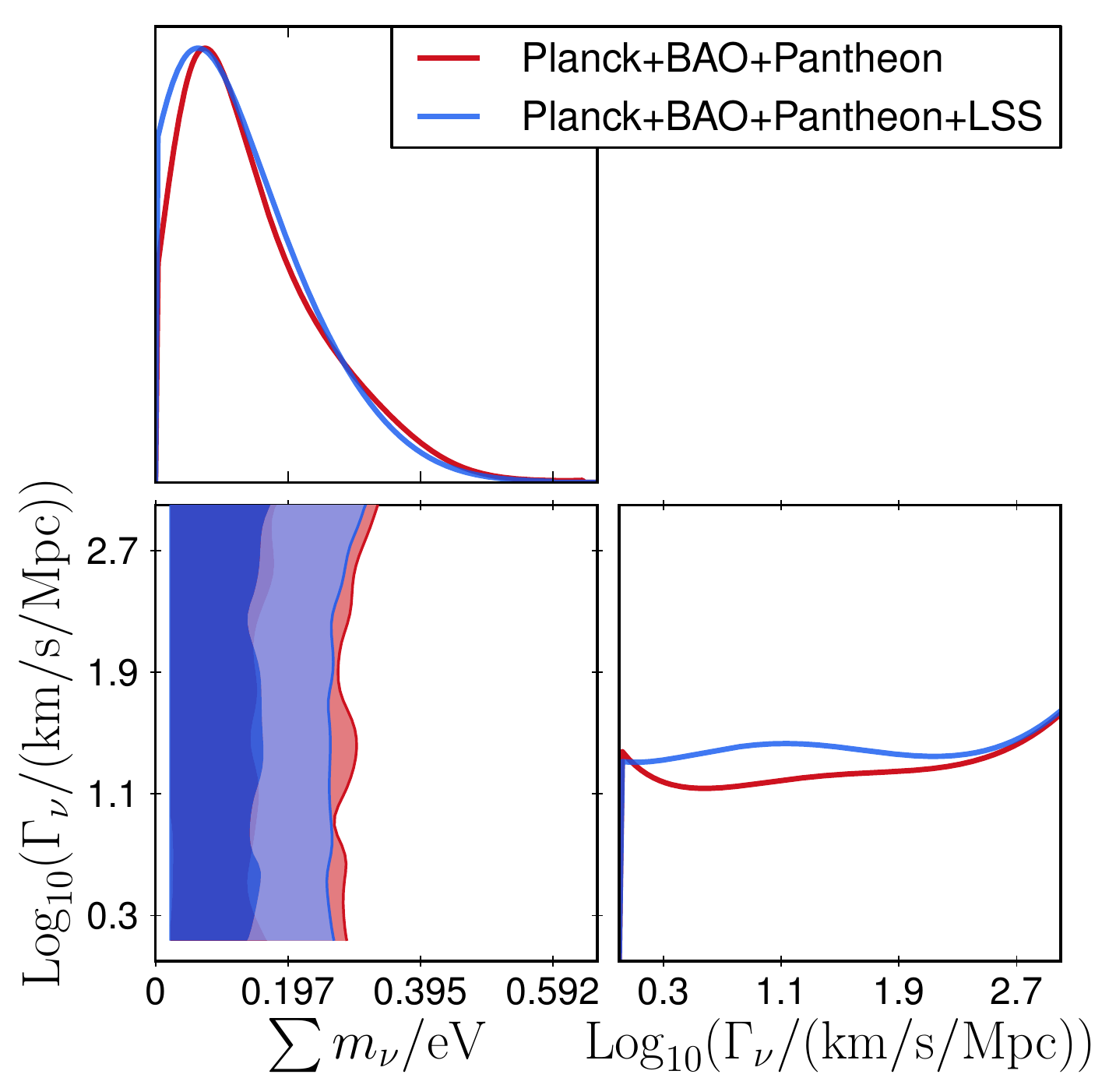}
     \includegraphics[scale=0.47]{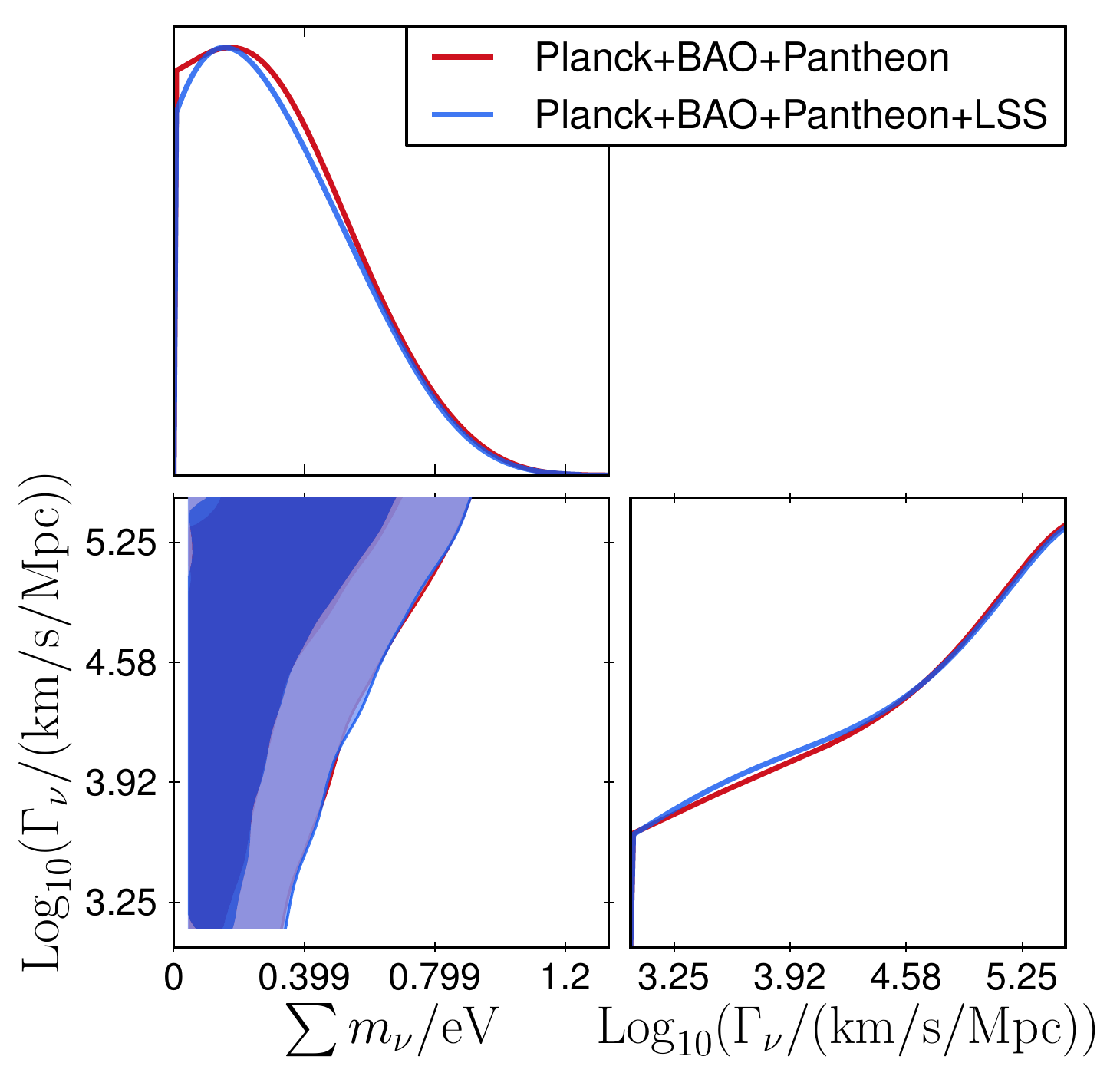}
    \caption{Posterior distributions of $\sum m_\nu$ and ${\rm log}_{10}\Gamma_\nu$ for each dataset. Small decay rate  ${\rm log}_{10}\Gamma_\nu/({\rm km/s/Mpc})\in[0,3]$ are shown in the left panel, while large decay rate  ${\rm log}_{10}\Gamma_\nu/({\rm km/s/Mpc})\in[3,5.5]$  are shown in the right panel.}
    \label{fig:current_constraints}
\end{figure}

In Fig.~\ref{fig:current_constraints} we show the 1D and 2D 
marginalized posterior distribution of ${\sum m_\nu}$ and ${\rm 
log}_{10}\Gamma_\nu$ for both datasets, cutting the parameter space 
between small decay rate ${\rm log}_{10}\Gamma_\nu/({\rm km/s/Mpc})\in[0,3]$ (left panel) 
and large decay rate ${\rm log}_{10}\Gamma_\nu/({\rm km/s/Mpc})\in[3,5.5]$ (right panel) to accelerate convergence.
Strikingly, once the neutrino lifetime is let free to vary, the 
constraint on $\sum m_\nu$ is driven by our prior on ${\rm 
log}_{10}\Gamma_\nu$. We recall that this was chosen in order to ensure 
that neutrinos decay while non-relativistic. Interestingly, the 
constraint stays quite stable for ${\rm log}_{10}\Gamma_\nu/({\rm km/s/Mpc})<2.5$, but 
relaxes to $\sum m_\nu<0.9$ eV (with Planck+BAO+Growth Function+Pantheon) for higher values of the 
decay rate. We note that the limit only marginally improves with the addition of 
current LSS data, especially at high decay rates (right panel), for which the improvement is below numerical noise.

Our study allows us to obtain a bound on the sum of neutrino masses as a 
function of the neutrino lifetime. We see that $\sum m_\nu $ can be as 
large as 0.90 eV for neutrinos that decay close to recombination. 
However, given our restricted prior enforcing non-relativistic decays, 
our analysis does not set a true upper bound on the neutrino mass. In 
order to derive the true upper bound we would need to correctly 
incorporate relativistic decays, taking into account inverse decay 
processes. We refer to Ref.~\cite{Archidiacono:2013dua,Escudero:2019gfk} for a discussion 
of that regime, and defer to future work a reanalysis of that region of 
parameter space in light of the latest {\em Planck} results.

\section{Conclusions}\label{sec:conclusion}

The fact that the couplings of neutrinos to the other SM particles are 
so weak makes it extremely difficult to study their properties. Even 
though it has been over six decades since neutrinos were first directly 
observed in the laboratory, several of their fundamental properties, 
including their masses and lifetimes, remain to be determined. However, 
neutrinos are also among the most abundant particles in the universe, 
and their gravitational pull has effects on cosmological observables.  
The universe is therefore an excellent laboratory for studying the 
detailed properties of neutrinos.

In this paper, we have explored the cosmological signals arising from 
the theoretically well-motivated scenario in which neutrinos decay into 
invisible dark radiation on timescales less than the age of the 
universe. We have studied the effects of neutrino decay on the evolution 
of density perturbations, both analytically and numerically, and used 
the results to generalize the bound on the sum of neutrino mass to the 
case when the lifetime of the neutrino is less than the age of the 
universe. We have shown that the existing mass bound from CMB and LSS 
measurements, which assumes that neutrinos are stable, gets weakened if 
neutrinos decay, so that values of $\sum m_\nu$ as large as 0.9 eV are 
still allowed by the data. This provides strong motivation to continue 
the current efforts to measure the neutrino masses directly in the lab, 
in spite of the limited reach of these experiments. Our analytical 
results show that the signals of neutrino decay in LSS and CMB-lensing 
primarily arise from the contributions of neutrinos and their daughters 
to the overall energy density, and are quite insensitive to their 
contributions to the fluctuations about the background. Although the bounds we obtain based on the existing data do not set independent constraints on the neutrino mass and lifetime, next generation measurements of the matter power spectrum at different redshifts will provide useful information that may help in breaking this degeneracy. We will explore this in the future work~\cite{Chacko:2020hmh}.

 \acknowledgments 
 We would like to thank warmly Thejs Brinckmann for his help with {\sc 
MontePython-v3}, and Marilena Loverde and Gustavo Marques-Tavares for 
useful discussions. ZC, AD, PD and YT are supported in part by the 
National Science Foundation under Grant Number PHY-1620074. ZC would 
like to thank the Fermilab Theory Group for hospitality during the 
completion of this work. ZC's stay at Fermilab was supported by the 
Fermilab Intensity Frontier Fellowship and by the Visiting Scholars 
Award \#17-S-02 from the Universities Research Association. VP thanks 
the Johns Hopkins University and particularly Marc Kamionkowski for his 
hospitality during completion of this work. This work has been made 
possible thanks to the facilities offered by the Universit\'{e} Savoie 
Mont-Blanc MUST computing center.

\appendix
\section{A Model of Massive Neutrino Decay into Dark Radiation}\label{app:1}

In this appendix we present a simple, realistic model in which massive 
neutrinos decay into invisible dark radiation on timescales of order the 
age of the universe. To illustrate the main features of the model, we 
first consider a simplified version with just a single flavor of SM 
neutrino, denoted by $\nu$, and two singlet right-handed neutrinos, 
labelled as $n$ and $n'$. The model also contains two complex scalars, 
labelled as $\Phi$ and $\Phi'$. We introduce $U(1)_n\times U(1)_{n'}$ 
global symmetries that act on the right-handed neutrinos. While $n$ and 
$\Phi$ carry equal and opposite charges under $U(1)_n$, $n'$ and $\Phi'$ 
are neutral under this symmetry. Similarly, $n'$ and $\Phi'$ carry equal 
and opposite charges under $U(1)_{n'}$, while $n$ and $\Phi$ are 
neutral. Then the part of the Lagrangian responsible for generating the 
neutrino masses takes the form,
 \bea\label{eq:L_1gen}
-\mathcal L \supset \frac{y}{\Lambda}\bar L \tilde H n \Phi+\frac{y'}{\Lambda}\bar L\tilde H n' \Phi'+\textrm{H.c.}
 \eea
 Here $L$ represents the SM lepton doublet and $\tilde H=i\sigma_2 H^*$, 
where $H$ denotes the SM Higgs doublet. $\Lambda$ is a UV mass scale 
while $y$ and $y'$ are coupling constants. Although this Lagrangian is 
nonrenormalizable, it can be interpreted as the low energy description 
of a renormalizable theory after particles with masses of $\Lambda$ have 
been integrated out. For example, consider the renormalizable Lagrangian,
 \bea
-\mathcal L= \tilde{y}\bar L \tilde H N + M_N N N^C + \tilde{\lambda} n N^c \Phi +
\tilde{y}'\bar L\tilde H N' + M'_N N' {N^{c}}' + \tilde{\lambda}' n' {N^{c}}' \Phi'
+\textrm{H.c.}
 \eea
 Terms of the form shown in Eqn.~(\ref{eq:L_1gen}) are obtained after 
the heavy fermions $N$, $N^c$ $N'$ and ${N^c}'$ have been integrated out.

Once the scalars $\Phi$, $\Phi'$ and the SM Higgs each acquire a vacuum 
expectation value (VEV), we obtain Dirac masses for the SM neutrino,
 \bea\label{eq:L_mass_1gen}
-\mathcal L&\supset&\frac{y f v}{2\Lambda}\bar\nu n +\frac{y' f' v}{2\Lambda}\bar\nu n' +\textrm{H.c.}\nonumber\\
&=&m \bar\nu n_{\rm h}+\textrm{H.c.}.
 \eea
 Here $\frac{f}{\sqrt{2}}$, $\frac{f'}{\sqrt{2}}$ and 
$\frac{v}{\sqrt{2}}$ denote the VEVs of $\Phi$, $\Phi'$ and $H$ 
respectively. The SM neutrino acquires a mass $m=\sqrt{\left(y 
f\right)^2+\left(y' f' \right)^2} v/(2\Lambda)$. Its Dirac partner 
$n_{\rm h}$ is one linear combination of $n$ and $n'$,
 \bea\label{eq:transformation}
\left(\begin{array}{c}
n_{\textrm{h}}\\
n_{\textrm{l}}
\end{array}\right)=
\left(\begin{array}{cc}
\cos \theta &\sin \theta \\
-\sin \theta &\cos \theta
\end{array}\right)
\left(\begin{array}{c}
n\\
n'
\end{array}\right)~;~\cos \theta=\frac{y f}{\sqrt{\left(y f \right)^2+\left(y' f' \right)^2}}.
 \eea
 It is clear from Eq.~(\ref{eq:L_mass_1gen}) that the spectrum contains 
one massive Dirac neutrino and one massless singlet neutrino $n_{\rm 
l}$.
 
Below the spontaneous symmetry breaking scales $f$ and $f'$, the Goldstone 
bosons can be parametrized as
 \bea
\Phi=\frac{f}{\sqrt{2}}e^{i \phi/f}~~,~~\Phi'=\frac{f'}{\sqrt{2}}e^{i \phi'/f'},
 \eea
 where $\phi$ and $\phi'$ denote the Goldstone bosons from $U(1)_n$ and 
$U(1)_{n'}$ respectively. The couplings of the Goldstone bosons in the 
low energy effective theory are dictated by the non-linearly realized 
global symmetries. To leading order in $1/f$ and $1/f'$, they are given 
by,
 \bea
-\mathcal L&\supset&i\frac{yf v}{2\Lambda}\frac{\phi}{f}\bar\nu n+i\frac{yf' v}{2\Lambda}\frac{\phi'}{f'}\bar\nu n +\textrm{H.c.}
 \eea 
 In the mass basis these interactions take the form,
 \bea
-\mathcal L&\supset&i m\bar\nu\left[\left(\frac{\phi}{f}\cos^2 \theta+\frac{\phi'}{f'}\sin^2 \theta\right)n_{\rm h}+\left(\frac{\phi'}{f'}-\frac{\phi}{f}\right)\sin\theta\cos\theta~n_{\rm l} \right]+\textrm{H.c.}
 \eea
 We see from this that the massive neutrino can decay into $n_{\rm l}$ 
and either $\phi$ or $\phi'$. Its partial widths into these decay modes 
are given by,
 \bea\label{eq:width_1gen}
\Gamma(\nu \to n_{\rm l} \phi)=\frac{m^3}{32\pi \bar f^2}~,~\Gamma(\nu \to n_{\rm l} \phi')=\frac{m^3}{32\pi \bar f^{\prime 2}},
 \eea
 where $\bar f\equiv f/(\cos \theta\sin\theta)$ and $\bar f'\equiv 
f'/(\cos \theta\sin\theta)$.

 Now we move on to discuss the realistic case in which there are three 
flavors of SM neutrinos $\nu_\alpha$ ($\alpha=e\,,\mu\,,\tau$). We also 
introduce three flavors of the sterile neutrinos $n_\alpha$ and 
$n'_\alpha$, as well as a new scalar field $\Sigma_{\alpha \beta}$. The 
global symmetry in the neutrino sector is now extended to $SU(3)_L\times 
SU(3)_R\times U(1)_n\times U(1)_{n'}$. The charge assignments under 
$U(1)_n\times U(1)_{n'} $ are the same as before, but with all 3 flavors 
of $n_\alpha$ and $n'_\alpha$ now being charged under $U(1)_n$ and $ 
U(1)_{n'}$ respectively. Under $SU(3)_L\times SU(3)_R$, the various 
fields transform as
 \bea
L \to U_L\,L \qquad n \to U_R\,n \qquad n' \to U_R\,n' \qquad \Sigma\to U_L\Sigma U_R^\dagger\,,
 \eea
 where $U_L$ and $U_R$ are the rotation matrices of $SU(3)_L$ and 
$SU(3)_R$ respectively. The neutrino masses now arise from terms in the 
Lagrangian of the form,
 \bea
-\mathcal L \supset \frac{y}{\Lambda^2}\bar L_\alpha \tilde H \Sigma_{\alpha \beta}n_\beta \Phi+\frac{y'}{\Lambda^2}\bar L_\alpha \tilde H\Sigma_{\alpha \beta}n'_\beta  \Phi'+\textrm{H.c.}
 \eea
 Once the $\Sigma$ field acquires a VEV, we can diagonalize its VEV 
$\langle\Sigma\rangle$ to obtain,
 \bea\label{eq:L_3gen}
-\mathcal L\supset\sum_i\left(\frac{y}{\Lambda^2}\bar L_i \tilde H \langle\Sigma\rangle_i n_i \Phi+\frac{y'}{\Lambda^2}\bar L_i \tilde H\langle\Sigma\rangle_i n'_i  \Phi'\right)+\textrm{H.c.}
 \eea
 where the index $i$ runs over $i=1,2,3$ and $\langle\Sigma\rangle_i$ 
denotes the $i$th eigenvalue of $\langle\Sigma\rangle$. The Lagrangian 
in Eq.~(\ref{eq:L_3gen}) can be viewed as three copies of 
Eq.~(\ref{eq:L_1gen}). After the scalars $\Phi$, $\Phi'$ and $H$ acquire 
VEVs, all three generations of $(n_i\,,n'_i)$ can be simultaneously 
transformed to the mass basis $(n_{\textrm{h}i},n_{\textrm{l}i})$ using 
the same orthogonal matrix,
 \bea
\left(\begin{array}{l}
n_{\textrm{h}i}\\
n_{\textrm{l}i}
\end{array}\right)=
\left(\begin{array}{cc}
\cos \theta &\sin \theta \\
-\sin \theta &\cos \theta
\end{array}\right)
\left(\begin{array}{l}
n_{i}\\
n'_{i}
\end{array}\right),
 \eea
 where $\cos\theta$ is exactly the same as in 
Eq.~(\ref{eq:transformation}). Now the neutrino masses are given by,
 \bea
m_i=\sqrt{\left(y f \right)^2+\left(y' f'\right)^2}\frac{\langle\Sigma\rangle_i v}{2\Lambda^2}.
 \eea
 Assuming that the Goldstone bosons from $\Sigma$ are heavier than the 
massive neutrinos due to some external source of explicit breaking, the 
dominant decay modes of the massive neutrinos are to a massless sterile 
neutrino and either $\phi$ or $\phi'$. Following the discussion above, 
the total neutrino decay width is given by
 \bea
\Gamma_{\nu_i} =\Gamma(\nu_i \to n_{\textrm{l}i} \phi)+\Gamma(\nu_i \to n_{\textrm{l}i} \phi')=\frac{m_i^3}{32\pi \bar f^2}+\frac{m_i^3}{32\pi \bar f^{\prime 2}},
 \eea
 where $\bar f$ and $\bar f'$ are as defined after 
Eq.~(\ref{eq:width_1gen}). One characteristic feature of this model is 
that the widths of the neutrinos scale as the cube of their masses, 
$\Gamma_{\nu_i}/\Gamma_{\nu_j}=m_i^3/m_j^3$. In the case of 
quasi-degenerate neutrinos, $m_1\approx m_2\approx m_3$, it is clear 
that all neutrinos have almost the same total width. Assuming $\bar 
f=\bar f'$, we find that the total width is of order $H_0$ for $\bar 
f\sim 10^5\, \GeV$ and neutrino masses of order 0.1 eV,
 \bea
\frac{\Gamma_{\nu_i}}{H_0}\approx  1.3 \left(\frac{m_i}{0.1 \; \textrm{eV}}\right)^3 \left(\frac{10^5 \; \GeV}{\bar f}\right)^2.
 \eea

 The parameter space of this model is constrained by astrophysical, 
cosmological and laboratory data. These limits are very similar to those 
on conventional Majoron models, and can be expressed in terms of bounds 
on the decay constants $f$ and $f'$. In the case of massless Goldstone 
bosons, the bounds from cosmology and astrophysics are the most severe. 
A strong cosmological constraint arises from requiring consistency with 
the observation that the cosmic neutrinos are free streaming at 
temperatures below an 
eV~\cite{Archidiacono:2013dua,Audren:2014lsa,Follin:2015hya,Escudero:2019gfk}. 
Neutrino-neutrino scattering mediated by Goldstone boson exchange can 
prevent the neutrinos from free streaming, impacting the heights and 
locations of the CMB peaks. This translates into constraints on $f$ and 
$f'$ of order 100 keV~\cite{Chacko:2003dt}. A stronger although somewhat 
model-dependent constraint, $f, f' \gtrsim 100$ MeV, may be obtained by 
requiring that the Goldstone bosons and right-handed neutrinos not 
contribute significantly to the energy density in radiation at the time 
of Big Bang nucleosynthesis (BBN), or during the CMB epoch.

The strongest astrophysical bounds arise from the effects of Goldstone 
bosons on supernovae. The large chemical potential for electron 
neutrinos inside the supernova means that these particles can now decay 
into final states containing a Goldstone boson and a right-handed 
neutrino. This has the effect of deleptonizing the core, preventing the 
explosion from taking place. In addition, the free streaming of 
Goldstone bosons out of the supernovae core can lead to overly rapid 
energy loss. The resulting constraints are at the level of $f, f' 
\gtrsim 100$ keV~\cite{Gelmini:1982rr, Kolb:1987qy, Choi:1989hi, 
Kachelriess:2000qc, Farzan:2002wx}. There are also bounds on the 
couplings of neutrinos to Goldstone bosons from laboratory experiments, 
such as neutrinoless double beta decay~\cite{Doi:1985dx,Doi:1987rx}, 
meson decays~\cite{Barger:1981vd, Gelmini:1982rr}, charged lepton 
decays~\cite{Lessa:2007up} and tritium decay~\cite{Arcadi:2018xdd}. 
These constraints arise from corrections to the energy spectrum of the 
visible final states due to Goldstone boson emission. However, in all 
these cases, the limits are weaker than astrophysical and cosmological 
bounds on massless Goldstone bosons. Clearly, our benchmark values of 
$f,f' \sim 10^5$ GeV are easily consistent with all current bounds.


\bibliography{DecayNu}

\providecommand{\href}[2]{#2}\begingroup\raggedright\begin{thebibliography}{100}

\bibitem{Aghanim:2018eyx}
{\bf Planck} Collaboration, N.~Aghanim et~al., {\it {Planck 2018 results. VI.
  Cosmological parameters}},  \href{http://arxiv.org/abs/1807.06209}{{\tt
  arXiv:1807.06209}}.

\bibitem{Bond:1980ha}
J.~R. Bond, G.~Efstathiou, and J.~Silk, {\it {Massive Neutrinos and the Large
  Scale Structure of the Universe}},  {\em Phys. Rev. Lett.} {\bf 45} (1980)
  1980--1984. [,61(1980)].

\bibitem{Hu:1997mj}
W.~Hu, D.~J. Eisenstein, and M.~Tegmark, {\it {Weighing neutrinos with galaxy
  surveys}},  {\em Phys. Rev. Lett.} {\bf 80} (1998) 5255--5258,
  [\href{http://arxiv.org/abs/astro-ph/9712057}{{\tt astro-ph/9712057}}].

\bibitem{Wong:2011ip}
Y.~Y.~Y. Wong, {\it {Neutrino mass in cosmology: status and prospects}},  {\em
  Ann. Rev. Nucl. Part. Sci.} {\bf 61} (2011) 69--98,
  [\href{http://arxiv.org/abs/1111.1436}{{\tt arXiv:1111.1436}}].

\bibitem{Lattanzi:2017ubx}
M.~Lattanzi and M.~Gerbino, {\it {Status of neutrino properties and future
  prospects - Cosmological and astrophysical constraints}},  {\em Front.in
  Phys.} {\bf 5} (2018) 70, [\href{http://arxiv.org/abs/1712.07109}{{\tt
  arXiv:1712.07109}}].

\bibitem{Lesgourgues:2018ncw}
J.~Lesgourgues, G.~Mangano, G.~Miele, and S.~Pastor, {\em {Neutrino
  Cosmology}}.
\newblock Cambridge University Press, 2018.

\bibitem{Tanabashi:2018oca}
{\bf Particle Data Group} Collaboration, M.~Tanabashi et~al., {\it {Review of
  Particle Physics}},  {\em Phys. Rev.} {\bf D98} (2018), no.~3 030001.

\bibitem{Serpico:2007pt}
P.~D. Serpico, {\it {Cosmological neutrino mass detection: The best probe of
  neutrino lifetime}},  {\em Phys. Rev. Lett.} {\bf 98} (2007) 171301,
  [\href{http://arxiv.org/abs/astro-ph/0701699}{{\tt astro-ph/0701699}}].

\bibitem{Serpico:2008zza}
P.~D. Serpico, {\it {Neutrinos and cosmology: a lifetime relationship}},  {\em
  J. Phys. Conf. Ser.} {\bf 173} (2009) 012018.

\bibitem{Beacom:2004yd}
J.~F. Beacom, N.~F. Bell, and S.~Dodelson, {\it {Neutrinoless universe}},  {\em
  Phys. Rev. Lett.} {\bf 93} (2004) 121302,
  [\href{http://arxiv.org/abs/astro-ph/0404585}{{\tt astro-ph/0404585}}].

\bibitem{Farzan:2015pca}
Y.~Farzan and S.~Hannestad, {\it {Neutrinos secretly converting to lighter
  particles to please both KATRIN and the cosmos}},  {\em JCAP} {\bf 1602}
  (2016), no.~02 058, [\href{http://arxiv.org/abs/1510.02201}{{\tt
  arXiv:1510.02201}}].

\bibitem{Petcov:1976ff}
S.~T. Petcov, {\it {The Processes mu $\rightarrow$ e Gamma, mu $\rightarrow$ e
  e anti-e, Neutrino' $\rightarrow$ Neutrino gamma in the Weinberg-Salam Model
  with Neutrino Mixing}},  {\em Sov. J. Nucl. Phys.} {\bf 25} (1977) 340.
  [Erratum: Yad. Fiz.25,1336(1977)].

\bibitem{Goldman:1977jx}
J.~T. Goldman and G.~J. Stephenson, Jr., {\it {Limits on the Mass of the
  Muon-neutrino in the Absence of Muon Lepton Number Conservation}},  {\em
  Phys. Rev.} {\bf D16} (1977) 2256.

\bibitem{Marciano:1977wx}
W.~J. Marciano and A.~I. Sanda, {\it {Exotic Decays of the Muon and Heavy
  Leptons in Gauge Theories}},  {\em Phys. Lett.} {\bf 67B} (1977) 303--305.

\bibitem{Lee:1977tib}
B.~W. Lee and R.~E. Shrock, {\it {Natural Suppression of Symmetry Violation in
  Gauge Theories: Muon - Lepton and Electron Lepton Number Nonconservation}},
  {\em Phys. Rev.} {\bf D16} (1977) 1444.

\bibitem{Pal:1981rm}
P.~B. Pal and L.~Wolfenstein, {\it {Radiative Decays of Massive Neutrinos}},
  {\em Phys. Rev.} {\bf D25} (1982) 766.

\bibitem{Mohapatra:1991ng}
R.~N. Mohapatra and P.~B. Pal, {\it {Massive neutrinos in physics and
  astrophysics}},  {\em World Sci. Lect. Notes Phys.} {\bf 41} (1991) 1--318.

\bibitem{1992pmn..book.....B}
F.~{Boehm} and P.~{Vogel}, {\em {Physics of Massive Neutrinos}}.
\newblock June, 1992.

\bibitem{Gelmini:1980re}
G.~B. Gelmini and M.~Roncadelli, {\it {Left-Handed Neutrino Mass Scale and
  Spontaneously Broken Lepton Number}},  {\em Phys. Lett.} {\bf 99B} (1981)
  411--415.

\bibitem{Chikashige_1981}
Y.~Chikashige, R.~Mohapatra, and R.~Peccei, {\it Are there real goldstone
  bosons associated with broken lepton number?},  {\em Physics Letters B} {\bf
  98} (Jan, 1981) 265--268.

\bibitem{Georgi:1981pg}
H.~M. Georgi, S.~L. Glashow, and S.~Nussinov, {\it {Unconventional Model of
  Neutrino Masses}},  {\em Nucl. Phys.} {\bf B193} (1981) 297--316.

\bibitem{VALLE198387}
J.~Valle, {\it Fast neutrino decay in horizontal majoron models},  {\em Physics
  Letters B} {\bf 131} (1983), no.~1 87 -- 90.

\bibitem{Gelmini:1983ea}
G.~B. Gelmini and J.~W.~F. Valle, {\it {Fast Invisible Neutrino Decays}},  {\em
  Phys. Lett.} {\bf 142B} (1984) 181--187.

\bibitem{Dvali:2016uhn}
G.~Dvali and L.~Funcke, {\it {Small neutrino masses from gravitational
  $\theta$-term}},  {\em Phys. Rev.} {\bf D93} (2016), no.~11 113002,
  [\href{http://arxiv.org/abs/1602.03191}{{\tt arXiv:1602.03191}}].

\bibitem{Funcke:2019grs}
L.~Funcke, G.~Raffelt, and E.~Vitagliano, {\it {Distinguishing Dirac and
  Majorana neutrinos by their gravi-majoron decays}},
  \href{http://arxiv.org/abs/1905.01264}{{\tt arXiv:1905.01264}}.

\bibitem{Bahcall:1972my}
J.~N. Bahcall, N.~Cabibbo, and A.~Yahil, {\it {Are neutrinos stable
  particles?}},  {\em Phys. Rev. Lett.} {\bf 28} (1972) 316--318. [,285(1972)].

\bibitem{Berezhiani:1991vk}
Z.~G. Berezhiani, G.~Fiorentini, M.~Moretti, and A.~Rossi, {\it {Fast neutrino
  decay and solar neutrino detectors}},  {\em Z. Phys.} {\bf C54} (1992)
  581--586.

\bibitem{Acker:1991ej}
A.~Acker, S.~Pakvasa, and J.~T. Pantaleone, {\it {Decaying Dirac neutrinos}},
  {\em Phys. Rev.} {\bf D45} (1992) 1--4.

\bibitem{Barger:1998xk}
V.~D. Barger, J.~G. Learned, S.~Pakvasa, and T.~J. Weiler, {\it {Neutrino decay
  as an explanation of atmospheric neutrino observations}},  {\em Phys. Rev.
  Lett.} {\bf 82} (1999) 2640--2643,
  [\href{http://arxiv.org/abs/astro-ph/9810121}{{\tt astro-ph/9810121}}].

\bibitem{Acker:1993sz}
A.~Acker and S.~Pakvasa, {\it {Solar neutrino decay}},  {\em Phys. Lett.} {\bf
  B320} (1994) 320--322, [\href{http://arxiv.org/abs/hep-ph/9310207}{{\tt
  hep-ph/9310207}}].

\bibitem{CHOUBEY200073}
S.~Choubey, S.~Goswami, and D.~Majumdar, {\it Status of the neutrino decay
  solution to the solar neutrino problem},  {\em Physics Letters B} {\bf 484}
  (2000), no.~1 73 -- 78.

\bibitem{Joshipura:2002fb}
A.~S. Joshipura, E.~Masso, and S.~Mohanty, {\it {Constraints on decay plus
  oscillation solutions of the solar neutrino problem}},  {\em Phys. Rev.} {\bf
  D66} (2002) 113008, [\href{http://arxiv.org/abs/hep-ph/0203181}{{\tt
  hep-ph/0203181}}].

\bibitem{Doroshkevich:1984}
A.~G. Doroshkevich and M.~Y. Khlopov, {\it {Formation of structure in a
  universe with unstable neutrinos}},  {\em Monthly Notices of the Royal
  Astronomical Society} {\bf 211} (11, 1984) 277--282.

\bibitem{Doroshkevich:1988}
A.~G. Doroshkevich, A.~A. Klyin, and M.~Y. Khlopov, {\it {Cosmological models
  with unstable neutrinos}},  {\em Soviet Astronomy} {\bf 32} (3, 1988)
  127--133.

\bibitem{Chianese:2018luo}
M.~Chianese, P.~Di~Bari, K.~Farrag, and R.~Samanta, {\it {Probing relic
  neutrino radiative decays with 21 cm cosmology}},  {\em Phys. Lett.} {\bf
  B790} (2019) 64--70, [\href{http://arxiv.org/abs/1805.11717}{{\tt
  arXiv:1805.11717}}].

\bibitem{Aalberts:2018obr}
J.~L. Aalberts et~al., {\it {Precision constraints on radiative neutrino decay
  with CMB spectral distortion}},  {\em Phys. Rev.} {\bf D98} (2018) 023001,
  [\href{http://arxiv.org/abs/1803.00588}{{\tt arXiv:1803.00588}}].

\bibitem{Beda:2013mta}
A.~G. Beda, V.~B. Brudanin, V.~G. Egorov, D.~V. Medvedev, V.~S. Pogosov, E.~A.
  Shevchik, M.~V. Shirchenko, A.~S. Starostin, and I.~V. Zhitnikov, {\it {Gemma
  experiment: The results of neutrino magnetic moment search}},  {\em Phys.
  Part. Nucl. Lett.} {\bf 10} (2013) 139--143.

\bibitem{Borexino:2017fbd}
{\bf Borexino} Collaboration, M.~Agostini et~al., {\it {Limiting neutrino
  magnetic moments with Borexino Phase-II solar neutrino data}},  {\em Phys.
  Rev.} {\bf D96} (2017), no.~9 091103,
  [\href{http://arxiv.org/abs/1707.09355}{{\tt arXiv:1707.09355}}].

\bibitem{Raffelt:1990pj}
G.~G. Raffelt, {\it {New bound on neutrino dipole moments from globular cluster
  stars}},  {\em Phys. Rev. Lett.} {\bf 64} (1990) 2856--2858.

\bibitem{Raffelt:1999gv}
G.~G. Raffelt, {\it {Limits on neutrino electromagnetic properties: An
  update}},  {\em Phys. Rept.} {\bf 320} (1999) 319--327.

\bibitem{Arceo-Diaz:2015pva}
S.~Arceo-D{\'\i}az, K.~P. Schr{\"o}der, K.~Zuber, and D.~Jack, {\it {Constraint
  on the magnetic dipole moment of neutrinos by the tip-RGB luminosity in
  $\omega$-Centauri}},  {\em Astropart. Phys.} {\bf 70} (2015) 1--11.

\bibitem{Peebles}
P.~J.~E. Peebles, {\it {The Role of Neutrinos in the Evolution of Primeval
  Adiabatic Perturbations}},  {\em Astrophys. J.} {\bf 180} (1973) 1.

\bibitem{Hu:1995en}
W.~Hu and N.~Sugiyama, {\it {Small scale cosmological perturbations: An
  Analytic approach}},  {\em Astrophys. J.} {\bf 471} (1996) 542--570,
  [\href{http://arxiv.org/abs/astro-ph/9510117}{{\tt astro-ph/9510117}}].

\bibitem{Bashinsky:2003tk}
S.~Bashinsky and U.~Seljak, {\it {Neutrino perturbations in CMB anisotropy and
  matter clustering}},  {\em Phys.Rev.} {\bf D69} (2004) 083002,
  [\href{http://arxiv.org/abs/astro-ph/0310198}{{\tt astro-ph/0310198}}].

\bibitem{Archidiacono:2013dua}
M.~Archidiacono and S.~Hannestad, {\it {Updated constraints on non-standard
  neutrino interactions from Planck}},  {\em JCAP} {\bf 1407} (2014) 046,
  [\href{http://arxiv.org/abs/1311.3873}{{\tt arXiv:1311.3873}}].

\bibitem{Audren:2014lsa}
B.~Audren et~al., {\it {Robustness of cosmic neutrino background detection in
  the cosmic microwave background}},  {\em JCAP} {\bf 1503} (2015) 036,
  [\href{http://arxiv.org/abs/1412.5948}{{\tt arXiv:1412.5948}}].

\bibitem{Follin:2015hya}
B.~Follin, L.~Knox, M.~Millea, and Z.~Pan, {\it {First Detection of the
  Acoustic Oscillation Phase Shift Expected from the Cosmic Neutrino
  Background}},  {\em Phys. Rev. Lett.} {\bf 115} (2015), no.~9 091301,
  [\href{http://arxiv.org/abs/1503.07863}{{\tt arXiv:1503.07863}}].

\bibitem{Escudero:2019gfk}
M.~Escudero and M.~Fairbairn, {\it {Cosmological Constraints on Invisible
  Neutrino Decays Revisited}},  \href{http://arxiv.org/abs/1907.05425}{{\tt
  arXiv:1907.05425}}.

\bibitem{Kreisch:2019yzn}
C.~D. Kreisch, F.-Y. Cyr-Racine, and O.~Dor{\'e}, {\it {The Neutrino Puzzle:
  Anomalies, Interactions, and Cosmological Tensions}},
  \href{http://arxiv.org/abs/1902.00534}{{\tt arXiv:1902.00534}}.

\bibitem{Frieman:1987as}
J.~A. Frieman, H.~E. Haber, and K.~Freese, {\it {Neutrino Mixing, Decays and
  Supernova Sn1987a}},  {\em Phys. Lett.} {\bf B200} (1988) 115--121.

\bibitem{Beacom:2002cb}
J.~F. Beacom and N.~F. Bell, {\it {Do solar neutrinos decay?}},  {\em Phys.
  Rev.} {\bf D65} (2002) 113009,
  [\href{http://arxiv.org/abs/hep-ph/0204111}{{\tt hep-ph/0204111}}].

\bibitem{Bandyopadhyay:2002qg}
A.~Bandyopadhyay, S.~Choubey, and S.~Goswami, {\it {Neutrino decay confronts
  the SNO data}},  {\em Phys. Lett.} {\bf B555} (2003) 33--42,
  [\href{http://arxiv.org/abs/hep-ph/0204173}{{\tt hep-ph/0204173}}].

\bibitem{GonzalezGarcia:2008ru}
M.~C. Gonzalez-Garcia and M.~Maltoni, {\it {Status of Oscillation plus Decay of
  Atmospheric and Long-Baseline Neutrinos}},  {\em Phys. Lett.} {\bf B663}
  (2008) 405--409, [\href{http://arxiv.org/abs/0802.3699}{{\tt
  arXiv:0802.3699}}].

\bibitem{Gomes:2014yua}
R.~A. Gomes, A.~L.~G. Gomes, and O.~L.~G. Peres, {\it {Constraints on neutrino
  decay lifetime using long-baseline charged and neutral current data}},  {\em
  Phys. Lett.} {\bf B740} (2015) 345--352,
  [\href{http://arxiv.org/abs/1407.5640}{{\tt arXiv:1407.5640}}].

\bibitem{Choubey:2018cfz}
S.~Choubey, D.~Dutta, and D.~Pramanik, {\it {Invisible neutrino decay in the
  light of NOvA and T2K data}},  {\em JHEP} {\bf 08} (2018) 141,
  [\href{http://arxiv.org/abs/1805.01848}{{\tt arXiv:1805.01848}}].

\bibitem{Aharmim:2018fme}
{\bf SNO} Collaboration, B.~Aharmim et~al., {\it {Constraints on Neutrino
  Lifetime from the Sudbury Neutrino Observatory}},  {\em Phys. Rev.} {\bf D99}
  (2019), no.~3 032013, [\href{http://arxiv.org/abs/1812.01088}{{\tt
  arXiv:1812.01088}}].

\bibitem{Blas:2011rf}
D.~Blas, J.~Lesgourgues, and T.~Tram, {\it {The Cosmic Linear Anisotropy
  Solving System (CLASS) II: Approximation schemes}},  {\em JCAP} {\bf 1107}
  (2011) 034, [\href{http://arxiv.org/abs/1104.2933}{{\tt arXiv:1104.2933}}].

\bibitem{Ade:2015xua}
{\bf Planck} Collaboration, P.~A.~R. Ade et~al., {\it {Planck 2015 results.
  XIII. Cosmological parameters}},  {\em Astron. Astrophys.} {\bf 594} (2016)
  A13, [\href{http://arxiv.org/abs/1502.01589}{{\tt arXiv:1502.01589}}].

\bibitem{Aghanim:2019ame}
{\bf Planck} Collaboration, N.~Aghanim et~al., {\it {Planck 2018 results. V.
  CMB power spectra and likelihoods}},
  \href{http://arxiv.org/abs/1907.12875}{{\tt arXiv:1907.12875}}.

\bibitem{Angrik:2005ep}
{\bf KATRIN} Collaboration, J.~Angrik et~al., {\it {KATRIN design report
  2004}}, .

\bibitem{KamLAND-Zen:2016pfg}
{\bf KamLAND-Zen} Collaboration, A.~Gando et~al., {\it {Search for Majorana
  Neutrinos near the Inverted Mass Hierarchy Region with KamLAND-Zen}},  {\em
  Phys. Rev. Lett.} {\bf 117} (2016), no.~8 082503,
  [\href{http://arxiv.org/abs/1605.02889}{{\tt arXiv:1605.02889}}]. [Addendum:
  Phys. Rev. Lett.117,no.10,109903(2016)].

\bibitem{Auger:2012gs}
M.~Auger et~al., {\it {The EXO-200 detector, part I: Detector design and
  construction}},  {\em JINST} {\bf 7} (2012) P05010,
  [\href{http://arxiv.org/abs/1202.2192}{{\tt arXiv:1202.2192}}].

\bibitem{Albert:2014awa}
{\bf EXO-200} Collaboration, J.~B. Albert et~al., {\it {Search for Majorana
  neutrinos with the first two years of EXO-200 data}},  {\em Nature} {\bf 510}
  (2014) 229--234, [\href{http://arxiv.org/abs/1402.6956}{{\tt
  arXiv:1402.6956}}].

\bibitem{Brunner:2017iql}
T.~Brunner and L.~Winslow, {\it {Searching for $0\nu\beta\beta$ decay in
  $^{136}$Xe -- towards the tonne-scale and beyond}},  {\em Nucl. Phys. News}
  {\bf 27} (2017), no.~3 14--19, [\href{http://arxiv.org/abs/1704.01528}{{\tt
  arXiv:1704.01528}}].

\bibitem{Audren:2014bca}
B.~Audren, J.~Lesgourgues, G.~Mangano, P.~D. Serpico, and T.~Tram, {\it
  {Strongest model-independent bound on the lifetime of Dark Matter}},  {\em
  JCAP} {\bf 1412} (2014), no.~12 028,
  [\href{http://arxiv.org/abs/1407.2418}{{\tt arXiv:1407.2418}}].

\bibitem{Poulin:2016nat}
V.~Poulin, P.~D. Serpico, and J.~Lesgourgues, {\it {A fresh look at linear
  cosmological constraints on a decaying dark matter component}},  {\em JCAP}
  {\bf 1608} (2016), no.~08 036, [\href{http://arxiv.org/abs/1606.02073}{{\tt
  arXiv:1606.02073}}].

\bibitem{Kawasaki:1992kg}
M.~Kawasaki, G.~Steigman, and H.-S. Kang, {\it {Cosmological evolution of an
  early decaying particle}},  {\em Nucl. Phys.} {\bf B403} (1993) 671--706.

\bibitem{Bharadwaj:1997dz}
S.~Bharadwaj and S.~K. Sethi, {\it {Decaying neutrinos and large scale
  structure formation}},  {\em Astrophys. J. Suppl.} {\bf 114} (1998) 37,
  [\href{http://arxiv.org/abs/astro-ph/9707143}{{\tt astro-ph/9707143}}].

\bibitem{Kaplinghat:1999xy}
M.~Kaplinghat, R.~E. Lopez, S.~Dodelson, and R.~J. Scherrer, {\it {Improved
  treatment of cosmic microwave background fluctuations induced by a late
  decaying massive neutrino}},  {\em Phys. Rev.} {\bf D60} (1999) 123508,
  [\href{http://arxiv.org/abs/astro-ph/9907388}{{\tt astro-ph/9907388}}].

\bibitem{Aoyama:2011ba}
S.~Aoyama, K.~Ichiki, D.~Nitta, and N.~Sugiyama, {\it {Formulation and
  constraints on decaying dark matter with finite mass daughter particles}},
  {\em JCAP} {\bf 1109} (2011) 025, [\href{http://arxiv.org/abs/1106.1984}{{\tt
  arXiv:1106.1984}}].

\bibitem{Wang:2012eka}
M.-Y. Wang and A.~R. Zentner, {\it {Effects of Unstable Dark Matter on
  Large-Scale Structure and Constraints from Future Surveys}},  {\em Phys.
  Rev.} {\bf D85} (2012) 043514, [\href{http://arxiv.org/abs/1201.2426}{{\tt
  arXiv:1201.2426}}].

\bibitem{Aoyama:2014tga}
S.~Aoyama, T.~Sekiguchi, K.~Ichiki, and N.~Sugiyama, {\it {Evolution of
  perturbations and cosmological constraints in decaying dark matter models
  with arbitrary decay mass products}},  {\em JCAP} {\bf 1407} (2014) 021,
  [\href{http://arxiv.org/abs/1402.2972}{{\tt arXiv:1402.2972}}].

\bibitem{Ma:1995ey}
C.-P. Ma and E.~Bertschinger, {\it {Cosmological perturbation theory in the
  synchronous and conformal Newtonian gauges}},  {\em Astrophys. J.} {\bf 455}
  (1995) 7--25, [\href{http://arxiv.org/abs/astro-ph/9506072}{{\tt
  astro-ph/9506072}}].

\bibitem{Lesgourgues13}
J.~Lesgourgues and T.~Tram, {\it {Fast and accurate CMB computations in
  non-flat FLRW universes}},  {\em JCAP} {\bf 1409} (2014), no.~09 032,
  [\href{http://arxiv.org/abs/1312.2697}{{\tt arXiv:1312.2697}}].

\bibitem{Archidiacono:2016lnv}
M.~Archidiacono, T.~Brinckmann, J.~Lesgourgues, and V.~Poulin, {\it {Physical
  effects involved in the measurements of neutrino masses with future
  cosmological data}},  {\em JCAP} {\bf 1702} (2017), no.~02 052,
  [\href{http://arxiv.org/abs/1610.09852}{{\tt arXiv:1610.09852}}].

\bibitem{Bernardeau:1996aa}
F.~Bernardeau, {\it {Weak lensing detection in CMB maps}},  {\em Astron.
  Astrophys.} {\bf 324} (1997) 15--26,
  [\href{http://arxiv.org/abs/astro-ph/9611012}{{\tt astro-ph/9611012}}].

\bibitem{1953ApJ...117..134L}
D.~N. {Limber}, {\it {The Analysis of Counts of the Extragalactic Nebulae in
  Terms of a Fluctuating Density Field.}},  {\em ApJ} {\bf 117} (Jan., 1953)
  134.

\bibitem{Pan:2014xua}
Z.~Pan, L.~Knox, and M.~White, {\it {Dependence of the Cosmic Microwave
  Background Lensing Power Spectrum on the Matter Density}},  {\em Mon. Not.
  Roy. Astron. Soc.} {\bf 445} (2014), no.~3 2941--2945,
  [\href{http://arxiv.org/abs/1406.5459}{{\tt arXiv:1406.5459}}].

\bibitem{Aghanim:2015xee}
{\bf Planck} Collaboration, N.~Aghanim et~al., {\it {Planck 2015 results. XI.
  CMB power spectra, likelihoods, and robustness of parameters}},  {\em Astron.
  Astrophys.} {\bf 594} (2016) A11,
  [\href{http://arxiv.org/abs/1507.02704}{{\tt arXiv:1507.02704}}].

\bibitem{Ade:2015zua}
{\bf Planck} Collaboration, P.~A.~R. Ade et~al., {\it {Planck 2015 results. XV.
  Gravitational lensing}},  {\em Astron. Astrophys.} {\bf 594} (2016) A15,
  [\href{http://arxiv.org/abs/1502.01591}{{\tt arXiv:1502.01591}}].

\bibitem{Beutler:2011hx}
F.~Beutler, C.~Blake, M.~Colless, D.~H. Jones, L.~Staveley-Smith, L.~Campbell,
  Q.~Parker, W.~Saunders, and F.~Watson, {\it {The 6dF Galaxy Survey: Baryon
  Acoustic Oscillations and the Local Hubble Constant}},  {\em Mon. Not. Roy.
  Astron. Soc.} {\bf 416} (2011) 3017--3032,
  [\href{http://arxiv.org/abs/1106.3366}{{\tt arXiv:1106.3366}}].

\bibitem{Ross:2014qpa}
A.~J. Ross, L.~Samushia, C.~Howlett, W.~J. Percival, A.~Burden, and M.~Manera,
  {\it {The clustering of the SDSS DR7 main Galaxy sample -- I. A 4 per cent
  distance measure at $z = 0.15$}},  {\em Mon. Not. Roy. Astron. Soc.} {\bf
  449} (2015), no.~1 835--847, [\href{http://arxiv.org/abs/1409.3242}{{\tt
  arXiv:1409.3242}}].

\bibitem{Alam:2016hwk}
{\bf BOSS} Collaboration, S.~Alam et~al., {\it {The clustering of galaxies in
  the completed SDSS-III Baryon Oscillation Spectroscopic Survey: cosmological
  analysis of the DR12 galaxy sample}},  {\em Mon. Not. Roy. Astron. Soc.} {\bf
  470} (2017), no.~3 2617--2652, [\href{http://arxiv.org/abs/1607.03155}{{\tt
  arXiv:1607.03155}}].

\bibitem{Scolnic:2017caz}
D.~M. Scolnic et~al., {\it {The Complete Light-curve Sample of
  Spectroscopically Confirmed SNe Ia from Pan-STARRS1 and Cosmological
  Constraints from the Combined Pantheon Sample}},  {\em Astrophys. J.} {\bf
  859} (2018), no.~2 101, [\href{http://arxiv.org/abs/1710.00845}{{\tt
  arXiv:1710.00845}}].

\bibitem{Reid:2009xm}
B.~A. Reid et~al., {\it {Cosmological Constraints from the Clustering of the
  Sloan Digital Sky Survey DR7 Luminous Red Galaxies}},  {\em Mon. Not. Roy.
  Astron. Soc.} {\bf 404} (2010) 60--85,
  [\href{http://arxiv.org/abs/0907.1659}{{\tt arXiv:0907.1659}}].

\bibitem{Kohlinger:2017sxk}
F.~K{\"o}hlinger et~al., {\it {KiDS-450: The tomographic weak lensing power
  spectrum and constraints on cosmological parameters}},  {\em Mon. Not. Roy.
  Astron. Soc.} {\bf 471} (2017), no.~4 4412--4435,
  [\href{http://arxiv.org/abs/1706.02892}{{\tt arXiv:1706.02892}}].

\bibitem{Audren:2012wb}
B.~Audren, J.~Lesgourgues, K.~Benabed, and S.~Prunet, {\it {Conservative
  Constraints on Early Cosmology: an illustration of the Monte Python
  cosmological parameter inference code}},  {\em JCAP} {\bf 1302} (2013) 001,
  [\href{http://arxiv.org/abs/1210.7183}{{\tt arXiv:1210.7183}}].

\bibitem{Brinckmann:2018cvx}
T.~Brinckmann and J.~Lesgourgues, {\it {MontePython 3: boosted MCMC sampler and
  other features}},  \href{http://arxiv.org/abs/1804.07261}{{\tt
  arXiv:1804.07261}}.

\bibitem{Lewis:2013hha}
A.~Lewis, {\it {Efficient sampling of fast and slow cosmological parameters}},
  {\em Phys. Rev.} {\bf D87} (2013), no.~10 103529,
  [\href{http://arxiv.org/abs/1304.4473}{{\tt arXiv:1304.4473}}].

\bibitem{Vagnozzi:2017ovm}
S.~Vagnozzi, E.~Giusarma, O.~Mena, K.~Freese, M.~Gerbino, S.~Ho, and
  M.~Lattanzi, {\it {Unveiling $\nu$ secrets with cosmological data: neutrino
  masses and mass hierarchy}},  {\em Phys. Rev.} {\bf D96} (2017), no.~12
  123503, [\href{http://arxiv.org/abs/1701.08172}{{\tt arXiv:1701.08172}}].

\bibitem{Chacko:2020hmh}
Z.~Chacko, A.~Dev, P.~Du, V.~Poulin, and Y.~Tsai, {\it {Determining the
  Neutrino Lifetime from Cosmology}},
  \href{http://arxiv.org/abs/2002.08401}{{\tt arXiv:2002.08401}}.

\bibitem{Chacko:2003dt}
Z.~Chacko, L.~J. Hall, T.~Okui, and S.~J. Oliver, {\it {CMB signals of neutrino
  mass generation}},  {\em Phys. Rev.} {\bf D70} (2004) 085008,
  [\href{http://arxiv.org/abs/hep-ph/0312267}{{\tt hep-ph/0312267}}].

\bibitem{Gelmini:1982rr}
G.~B. Gelmini, S.~Nussinov, and M.~Roncadelli, {\it {Bounds and Prospects for
  the Majoron Model of Left-handed Neutrino Masses}},  {\em Nucl. Phys.} {\bf
  B209} (1982) 157--173.

\bibitem{Kolb:1987qy}
E.~W. Kolb and M.~S. Turner, {\it {Supernova SN 1987a and the Secret
  Interactions of Neutrinos}},  {\em Phys. Rev.} {\bf D36} (1987) 2895.

\bibitem{Choi:1989hi}
K.~Choi and A.~Santamaria, {\it {Majorons and Supernova Cooling}},  {\em Phys.
  Rev.} {\bf D42} (1990) 293--306.

\bibitem{Kachelriess:2000qc}
M.~Kachelriess, R.~Tomas, and J.~W.~F. Valle, {\it {Supernova bounds on Majoron
  emitting decays of light neutrinos}},  {\em Phys. Rev.} {\bf D62} (2000)
  023004, [\href{http://arxiv.org/abs/hep-ph/0001039}{{\tt hep-ph/0001039}}].

\bibitem{Farzan:2002wx}
Y.~Farzan, {\it {Bounds on the coupling of the Majoron to light neutrinos from
  supernova cooling}},  {\em Phys. Rev.} {\bf D67} (2003) 073015,
  [\href{http://arxiv.org/abs/hep-ph/0211375}{{\tt hep-ph/0211375}}].

\bibitem{Doi:1985dx}
M.~Doi, T.~Kotani, and E.~Takasugi, {\it {Double beta Decay and Majorana
  Neutrino}},  {\em Prog. Theor. Phys. Suppl.} {\bf 83} (1985) 1.

\bibitem{Doi:1987rx}
M.~Doi, T.~Kotani, and E.~Takasugi, {\it {The Neutrinoless Double Beta Decay
  With Majoron Emission}},  {\em Phys. Rev.} {\bf D37} (1988) 2575.

\bibitem{Barger:1981vd}
V.~D. Barger, W.-Y. Keung, and S.~Pakvasa, {\it {Majoron Emission by
  Neutrinos}},  {\em Phys. Rev.} {\bf D25} (1982) 907.

\bibitem{Lessa:2007up}
A.~P. Lessa and O.~L.~G. Peres, {\it {Revising limits on neutrino-Majoron
  couplings}},  {\em Phys. Rev.} {\bf D75} (2007) 094001,
  [\href{http://arxiv.org/abs/hep-ph/0701068}{{\tt hep-ph/0701068}}].

\bibitem{Arcadi:2018xdd}
G.~Arcadi, J.~Heeck, F.~Heizmann, S.~Mertens, F.~S. Queiroz, W.~Rodejohann,
  M.~Slezák, and K.~Valerius, {\it {Tritium beta decay with additional
  emission of new light bosons}},  {\em JHEP} {\bf 01} (2019) 206,
  [\href{http://arxiv.org/abs/1811.03530}{{\tt arXiv:1811.03530}}].

\end{thebibliography}\endgroup
\bibliographystyle{JHEP}


\end{document}